\documentclass[aps,prl,10pt,twocolumn,superscriptaddress,floatfix,longbibliography]{revtex4-2} \usepackage{amsmath,amssymb,bm} \usepackage{graphicx} \usepackage{xcolor} \usepackage[colorlinks=true,allcolors=blue]{hyperref} \newcommand{\ii}{\mathrm{i}} \newcommand{\dd}{\mathrm{d}} \newcommand{\ket}[1]{\lvert#1\rangle} \newcommand{\bra}[1]{\langle#1\rvert}

\graphicspath{{arxiv/}}
\begin{document}

\title{Anyon Chern bands of opposite sign and reversed drift}

\author{Zhenyu Xiao}
\email{zyxiao@princeton.edu}
\affiliation{Princeton Quantum Initiative, Princeton University, Princeton, New Jersey 08544, USA}
\affiliation{Princeton Center for Theoretical Science, Princeton University, Princeton, New Jersey 08544, USA}

\author{Minxuan Wang}
\affiliation{Department of Physics, Princeton University, Princeton, New Jersey 08544, USA}

\author{Shinsei Ryu}
\affiliation{Department of Physics, Princeton University, Princeton, New Jersey 08544, USA}

\begin{abstract}
The elementary charged excitations of fractional quantum Hall states are anyons.
Without continuous translation symmetry, they acquire a dispersion and richer dynamics.
We study the Laughlin quasihole at filling $1/3$ in a periodically modulated magnetic field, a common model of fractional Chern insulators in two-dimensional materials.
From the microscopic wave functions, we derive the commutator of its guiding-center coordinates, set by a many-body Berry curvature that governs its wave packets.
A modulation with two flux quanta per unit cell splits the quasihole band into bands with Chern numbers $6$ and $-3$; such opposite signs are generic as long as a weak modulation opens a gap between the anyon bands.
In the $C=-3$ band, the quasihole drifts against $\bm E\times\bm B$, although the magnetic field has the same sign everywhere.
Perturbation theory in the modulation explains these results, and large-scale Monte Carlo calculations confirm them.
Anyon bands thus acquire their own topology, which controls anyon motion and may shape the phases of itinerant anyons.
\end{abstract}

\maketitle

\emph{Introduction.}---Anyons are quasiparticles with exchange statistics beyond those of bosons and fermions~\cite{Leinaas1977,Wilczek1982,Arovas1984,Stern2008}.
In fractional quantum Hall (FQH) states~\cite{Tsui1982,Laughlin1983}, they emerge as elementary charged excitations carrying fractional charge~\cite{Saminadayar1997,dePicciotto1997}.
Understanding how they move and couple to external fields is important both for many-body physics and for topological quantum computation~\cite{Kitaev2003,Nayak2008}.
Their dynamics~\cite{Bartolomei2020,Nakamura2020,Ruelle2025} also offers a window into the underlying topological order.

Fractional Chern insulator states~\cite{Tang2011,Sun2011,Neupert2011,Sheng2011,Regnault2011,Qi2011,Wu2012a,Bernevig2012,Wu2012b,Wu2013,Liu2015,Ledwith2020,Repellin2020,Ledwith2022,Dong2023b,Parker2026,Bernevig2025}
in twisted bilayer MoTe$_2$~\cite{Wu2019,Li2021,Devakul2021,Crepel2023,Reddy2023,Dong2023a,Cai2023,Zeng2023,Park2023,Xu2023,Wang2024,Yu2024,Ji2024,Redekop2024,Park2025HighT} and rhombohedral graphene~\cite{Zhang2019,Abouelkomsan2020,Lu2024,Dong2024a,Dong2024b,Zhou2024,Tan2024,Kwan2025,Xie2025,Aronson2025,Li2026c} provide new platforms for exploring anyons~\cite{Li2026b,Goncalves2026,Liu2025,Schleith2025,Yan2026,Li2026a,Wagner2026}.
The lattice breaks the continuous magnetic translation symmetry of a Landau level~\cite{Haldane1985b}, an effect that a periodically modulated magnetic field~\cite{Ledwith2020,Dong2022,Gao2023,MoralesDuran2024,Shi2024,Shi2026,Yan2026,Pu2026} captures through its connection to ideal Chern bands~\cite{Roy2014,Claassen2015,Wang2021,MeraOzawa2021,Ledwith2023,Estienne2023,Li2025,Yu_2025_WL_ideal,Zhao2026IdealBands}.
The absence of continuous magnetic translation symmetry leads to richer physics than in FQH systems: anyons acquire a dispersion~\cite{Goncalves2026,Liu2025,Iyer2026} and, once itinerant, are proposed to realize superconductors~\cite{Laughlin1988,Fetter1989,LeeFisher1989,Chen1989,WenZee1990,TangWen2013,Kim2025,Shi2025,Divic2025,ShiDelocalization2025,ShiNonAbelian2025,Nosov2026,Pichler2026,Lotric2026Itinerant,Fan2026WeakPairing,ShiSenthil2026,Mehta2026Coloring,Seo2026Wavefunctions,WangZaletel2025} and fractionalized metals~\cite{Shi2025,ShiSenthil2026,Senthil2026,Han2026Orthogonal,Zhang2026Color}.
Discrete translation symmetry also affects other collective modes~\cite{Repellin2014,Shen2026,Kousa2025,Wu2016,Wolf2025,Paul2026}.

A starting point for understanding these phases is the dynamics of a single anyon and its band structure.
An electron wave packet moves according to the dispersion and the Berry curvature of its band~\cite{ChangNiu1995,Sundaram1999,Xiao2010}.
The anyon dispersion has been studied extensively~\cite{Goncalves2026,Liu2025,Schleith2025,Yan2026,Iyer2026,Wang2026}, but whether a Berry curvature also governs anyon motion, and what this curvature is for a many-body Bloch state, remains unclear.
It would also characterize the topology of anyon bands, which may bear on the collective phases of itinerant anyons~\cite{TangWen2013,Shi2025,Divic2025,ShiDelocalization2025,ShiNonAbelian2025,Nosov2026,Pichler2026,Lotric2026Itinerant,Fan2026WeakPairing,ShiSenthil2026,Mehta2026Coloring,Seo2026Wavefunctions,Senthil2026,Han2026Orthogonal,Zhang2026Color,Han2026Superfluidity,Zhang2026Holon,Shi2025Melting,Wang2026Trilayer,Zhang2026Charge4e,Shi2026Parafermions,Zhang2026Transition,Shi2026Bogoliubov,Shi2026Metallogenic}.
One can then ask how much of this curvature and topology is inherited from the uniform-field FQH state and fixed by the ground-state topological order, and how much is new and tunable.

In this Letter, we study a single anyon, the Laughlin quasihole at filling $1/3$, in a periodically modulated magnetic field.
The interplay between interactions and modulation not only gives the anyon a dispersion but also modifies the commutator of its guiding-center coordinates, which is given by a many-body Berry curvature defined from the density form factor of the anyon Bloch states.
This Berry curvature governs the wave-packet dynamics of the anyon in an electric field and integrates to a Chern number $C=3$ for the anyon band, compared with $C=-1$ for the electronic Landau level.
We further engineer the anyon bands with a modulation of two flux quanta per unit cell, which splits the parent band with $C=3$ into bands with $C=6$ and $-3$; such opposite signs are generic whenever a weak modulation opens a gap~\cite{SM}.
As an immediate consequence, an anyon wave packet in the band with $C=-3$ drifts against the usual $\bm E\times\bm B$ direction, even though the magnetic field has the same sign everywhere.
We demonstrate this reversed drift by direct simulation of the wave-packet dynamics.
The dispersions and Berry curvatures are obtained analytically by perturbation theory in the modulation strength and confirmed by large-scale Monte Carlo calculations.

\emph{Anyon Bloch states with modulated magnetic field.}---We consider the Laughlin state at filling $1/3$ with one quasihole.
On a torus of size $L_x\times L_y$ in the uniform magnetic field $-B_0\hat{\bm z}$, with $N_e$ electrons and $N_\phi=3N_e+1$ flux quanta, its wave function is~\cite{Haldane1985a,Greiter2016}
\begin{equation}
 \begin{aligned}
 \psi^\sigma_\eta={}&e^{(\eta^2-|\eta|^2)/(12\ell_B^2)}\,
 e^{\sum_i(z_i^2-|z_i|^2)/(4\ell_B^2)}\,F^\sigma(Z+\tfrac\eta3)\\
 &\times\prod_i\vartheta_1\!\Big(\frac{z_i-\eta}{L_x}\,\Big|\,
 \frac{\ii L_y}{L_x}\Big)
 \prod_{i<j}\vartheta_1\!\Big(\frac{z_i-z_j}{L_x}\,\Big|\,
 \frac{\ii L_y}{L_x}\Big)^3
 \end{aligned}
 \label{eq:torus-qh}
\end{equation}
with the position of the quasihole $\eta$, $z_i=x_i+\ii y_i$, $Z=\sum_iz_i$, magnetic length $\ell_B=\sqrt{\hbar/eB_0}$, Jacobi theta function $\vartheta_1$ and center-of-mass factor $F^\sigma$ ($\sigma=0,1,2$)~\cite{SM}.
It is the exact zero mode of the Trugman--Kivelson (TK) interaction $V_{\rm TK}$~\cite{Trugman1985}.
While $\eta$ can move continuously on the torus and $\sigma = 0,1,2$ labels three topological sectors, the quasihole Hilbert space $\mathcal H_{\rm qh}$ these states span has a dimension $N_\phi$~\cite{Regnault2011,Fayyazuddin1996}.

A natural basis of $\mathcal H_{\rm qh}$ consists of the anyon Bloch states (i.e., many-body eigenstates of the total momentum of the electrons).
We divide the torus into $N_1\times N_2$ square cells with the primitive vectors $\bm a_1=a\hat{\bm x}$ and $\bm a_2=a\hat{\bm y}$, $a=\sqrt{2\pi}\ell_B$, each enclosing one flux quantum.
The magnetic translation $U_{\bm c}$ moves every electron by $\bm c$, $U_{\bm c}\psi(\{z_i\}) =e^{\ii\operatorname{Im}(\bar cZ)/(2\ell_B^2)}\,\psi(\{z_i-c\})$ with $c=c_x+\ii c_y$.
The translations $U_{\bm a_1}$ and $U_{\bm a_2}$ commute, and
\begin{equation}
 \ket{\Psi_{\bm k}}\propto\sum_{r=0}^{N_1-1}\sum_{t=0}^{N_2-1}
 e^{\ii\bm k\cdot(r\bm a_1+t\bm a_2)}\,
 U_{\bm a_1}^rU_{\bm a_2}^t\ket{\psi^\sigma_\eta}
 \label{eq:bloch}
\end{equation}
is their eigenstate with the total crystal momentum $\bm k$, $U_{\bm a_j}\ket{\Psi_{\bm k}}=e^{-\ii\bm k\cdot\bm a_j}\ket{\Psi_{\bm k}}$, the phase of a plane wave $e^{\ii\bm k\cdot\bm r}$ displaced by $\bm a_j$.
Since $U_{\bm a_j}^{N_j}=1$ on the torus, the momentum takes the $N_\phi=N_1N_2$ values $\bm k=(2\pi/a)(n_1/N_1,n_2/N_2)$, with $n_j=0,\ldots,N_j-1$.
Whatever $\eta$ and $\sigma$ one starts from, Eq.~\eqref{eq:bloch} gives the same state up to a factor, so that there is exactly one Bloch state at each momentum~\cite{Haldane1985b,SM}; we normalize the norm of $\ket{\Psi_{\bm k}}$ to unity.

We now modulate the field, $B(\bm r)=-B_0+\delta B(\bm r)$, where $\delta B$ is periodic with zero average.
For electrons with a $g$ factor of $2$, the lowest Landau level remains an exactly degenerate band of zero modes in the modulated field, the Aharonov--Casher (AC) band~\cite{Aharonov1979}.
The quasihole states in the AC band are $\Gamma\psi^\sigma_\eta$~\cite{Yan2026}, with
\begin{equation}
 \Gamma=\exp\Big[-\frac12\sum_iw(\bm r_i)\Big],
 \qquad \nabla^2w=-\frac{2e}{\hbar}\,\delta B.
 \label{eq:deformed}
\end{equation}
Since $\Gamma$ does not move the zeros of the wave function, they are still exact zero modes of $V_{\rm TK}$, and the quasihole space becomes $\Gamma\mathcal H_{\rm qh}$.
If the modulation keeps the translation symmetry of the cell, $\Gamma\ket{\Psi_{\bm k}}$ is still a Bloch state with the momentum $\bm k$.

\emph{Anyon dispersion, Berry curvature, and dynamics.}---We take the interaction $\alpha V_{\rm TK}+V$, and for concreteness $V$ is the Coulomb interaction screened by two gates at the distance $d$, $V(q)=(2\pi e^2/\epsilon q)\tanh(qd)$.
We consider large $\alpha$, so that we only need to consider the quasihole space $\Gamma\mathcal H_{\rm qh}$.
Without modulation, magnetic translation symmetry keeps the quasihole band flat under generic interactions~\cite{Haldane1985b}.
The quasihole carries fractional charge $e^*=e/3$.
Its low-energy dynamics can be described by guiding-center coordinates satisfying $[\mathcal R_x,\mathcal R_y]=\ii\hbar/e^*B_0=3\ii\ell_B^2$~\cite{Arovas1984,Kjonsberg1997}.
A weak uniform electric field $\bm E$ therefore produces the Hall drift $\dot{\bm{\mathcal R}}=\bm E\times\bm B/B_0^2$, with $\bm B=-B_0\hat{\bm z}$.

We then modulate the field without enlarging the unit cell: $\delta B$ has the periods $\bm a_1$ and $\bm a_2$, with one flux quantum per cell.
Since $\Gamma$ and $V$ commute with the lattice translations, so does $\mathcal P_{\rm qh}V\mathcal P_{\rm qh}$, where $\mathcal P_{\rm qh}$ projects onto $\Gamma\mathcal H_{\rm qh}$.
Each momentum sector contains only $\Gamma\ket{\Psi_{\bm k}}$, so that these states are the eigenstates, and they form a single band~\cite{Goncalves2026,Liu2025,Schleith2025,Yan2026,Iyer2026} with the energy
\begin{equation}
 E(\bm k)=\frac{\bra{\Psi_{\bm k}}\Gamma V\Gamma\ket{\Psi_{\bm k}}}
 {\mathcal N(\bm k)},\qquad
 \mathcal N(\bm k)=\bra{\Psi_{\bm k}}\Gamma^2\ket{\Psi_{\bm k}}.
 \label{eq:single-energy}
\end{equation}
This energy has been computed by Monte Carlo sampling of $\Gamma\ket{\Psi_{\bm k}}$~\cite{Schleith2025,Yan2026,Iyer2026} and by exact diagonalization of the projected interaction~\cite{Goncalves2026,Liu2025}.
For the first-harmonic modulation $\delta B=-B_0[\kappa_x\cos(\bm b_1\cdot\bm r)+\kappa_y\cos(\bm b_2\cdot\bm r)]$, with $\bm b_1=(2\pi/a)(1,0)$ and $\bm b_2=(2\pi/a)(0,1)$, it is to first order [\hyperref[sec:endmatter-single]{End Matter}]
\begin{equation}
 E(\bm k)-E_0=\frac{\chi_1}{\pi}
 [\kappa_y\cos(3k_xa)+\kappa_x\cos(3k_ya)],
 \label{eq:single-dispersion}
\end{equation}
where $E_0=\bra{\Psi_{\bm 0}}V\ket{\Psi_{\bm 0}}$ is the energy in the uniform field and $\chi_1=\bra{\Psi_{\bm 0}}(V-E_0)\rho_{\bm b_1}\ket{\Psi_{\bm 0}}$, with $\rho_{\bm q}=\sum_i e^{\ii\bm q\cdot\bm r_i}$; both are properties of the uniform state.

The modulation changes the commutator $[\mathcal R_x,\mathcal R_y]$.
We define the position of the quasihole as the center of its excess charge,
\begin{equation}
 \bm{\mathcal R}=\frac1{n_{\rm ex}}\,
 \mathcal P_{\rm qh}\!\int\!\dd^2r\,\bm r\,
 \big[\hat n(\bm r)-n_{\rm bg}(\bm r)\big]\mathcal P_{\rm qh}.
 \label{eq:position-def}
\end{equation}
Here $\hat n(\bm r)=\sum_i\delta(\bm r-\bm r_i)$ is the electron density, $n_{\rm bg}(\bm r)$ is the electron density of the $\nu=1/3$ ground state in the same magnetic field, which is the density far from the quasihole, and $n_{\rm ex}=-1/3$ is the excess electron number of the quasihole.
With modulation, $n_{\rm bg}$ is generally periodic but no longer uniform.
Consider a wave packet $\ket\zeta=\sum_{\bm k}\zeta(\bm k)\ket{\Psi^\Gamma_{\bm k}}$ of the normalized Bloch states $\ket{\Psi^\Gamma_{\bm k}}=[\mathcal N(\bm k)]^{-1/2}\Gamma\ket{\Psi_{\bm k}}$, with the momentum-space wave function $\zeta(\bm k)$.
In this basis, the position operator [Eq.~\eqref{eq:position-def}] acts as (see the \hyperref[sec:endmatter-dynamics]{End Matter})
\begin{equation}
 \bm{\mathcal R}=\ii\nabla_{\bm k}+\bm{\mathcal A}(\bm k),\qquad
 [\mathcal R_x,\mathcal R_y]=\ii\Omega(\bm k).
 \label{eq:position}
\end{equation}
The commutator is now set by the many-body Berry curvature $\Omega(\bm k)$, which depends on the momentum.
It is defined by the form factor of the density between Bloch states~\cite{Geraedts2018,Pu2026},
\begin{equation}
 \Lambda_{\bm q}(\bm k)=\bra{\Psi^\Gamma_{\bm k+\bm q}}\rho_{\bm q}
 \ket{\Psi^\Gamma_{\bm k}},\qquad
 \rho_{\bm q}=\sum_ie^{\ii\bm q\cdot\bm r_i},
 \label{eq:form-factor}
\end{equation}
as the phase of $\Lambda$ 
around a plaquette of the momentum mesh,
\begin{equation}
 \begin{aligned}
 &\Omega(\bm k)=\frac1{|\bm q_x||\bm q_y|}\operatorname{Arg}\big[
 \Lambda_{\bm q_x}(\bm k)\,\Lambda_{\bm q_y}(\bm k+\bm q_x)\\
 &\quad\times\Lambda_{-\bm q_x}(\bm k+\bm q_x+\bm q_y)\,
 \Lambda_{-\bm q_y}(\bm k+\bm q_y)\big],
 \end{aligned}
 \label{eq:plaquette}
\end{equation}
where $\bm q_x=(2\pi/L_x)\hat{\bm x}$ and $\bm q_y=(2\pi/L_y)\hat{\bm y}$ are the smallest momentum transfers on the torus.
The connection $\bm{\mathcal A}(\bm k)$, with $\Omega=\partial_{k_x}\mathcal A_y-\partial_{k_y}\mathcal A_x$, is the phase of the form factor between nearby momenta, $\Lambda_{\bm q}(\bm k)=n_{\rm ex}[1+\ii\bm q\cdot\bm{\mathcal A}(\bm k)+O(q^2)]$ at small $\bm q\neq0$.
This curvature describes the momentum-space geometry of the Bloch band, distinct from the real-space Berry curvature of localized quasiholes~\cite{Arovas1984,Yan2026}; a momentum-space Berry curvature has also been presented alongside $\mathcal N(\bm k)$~\cite{Yan2026}.

In the thermodynamic limit, and provided that the plasma that describes the modulated Laughlin state screens perfectly, we find that~\cite{SM}
\begin{equation}
 \Omega(\bm k)=3\ell_B^2-\frac12\nabla_{\bm k}^2
 \ln\mathcal N(\bm k).
 \label{eq:single-curvature}
\end{equation}
Recall that an ideal single-particle Chern band obtained from the lowest Landau level by a momentum-independent deformation~\cite{Ledwith2020,Wang2021} has the Berry curvature $-\ell_B^2-\frac12\nabla_{\bm k}^2\ln\mathcal N_{\rm id}$ in our convention, with $\mathcal N_{\rm id}$ the squared norm of the deformed Bloch state; the quasihole curvature has the same form, with $3\ell_B^2$ in place of $-\ell_B^2$.
For the first-harmonic modulation above, the curvature is to first order [\hyperref[sec:endmatter-single]{End Matter}]
\begin{equation}
 \Omega(\bm k)=3\ell_B^2\big\{1+\xi_1
 [\kappa_y\cos(3k_xa)+\kappa_x\cos(3k_ya)]\big\},
 \label{eq:single-curvature-first}
\end{equation}
with $\xi_1=3\bra{\Psi_{\bm 0}}\rho_{\bm b_1}\ket{\Psi_{\bm 0}}$, a property of the uniform state that does not depend on the interaction, in agreement with Monte Carlo (Fig.~\ref{fig:single-band}).

The Chern number of the band is $C=(2\pi)^{-1}\int\dd^2k\,\Omega(\bm k)$, integrated over the Brillouin zone of area $2\pi/\ell_B^2$.
On a finite torus it is the sum of the plaquette phases of Eq.~\eqref{eq:plaquette} divided by $2\pi$, an integer by construction.
Since $\mathcal N(\bm k)$ is periodic, the second term of Eq.~\eqref{eq:single-curvature} integrates to zero, and $C=3$ with or without modulation.
An ordinary particle of charge $e/3$ would have a Landau level with $C=1$: the curvature is the same, $3\ell_B^2$, but its magnetic length is $\sqrt3\,\ell_B$, so that its Brillouin zone has one third of the area.
The three topological sectors $\sigma$ on the torus triple the number of quasihole states, so that the Brillouin zone is that of the electrons and $C=3$.
The corresponding result for Jain states at $\nu=n/(2pn+1)$ is $|C|=2pn+1$~\cite{Pu2026}.
In our convention the lowest Landau level of the electrons has $C=-1$; the opposite sign reflects the charge of the quasihole, which is opposite to that of the electron.
This Chern number is not the Niu--Thouless--Wu invariant defined by twisted boundary conditions~\cite{Niu1985}, and it does not directly give the Hall conductance.
It governs instead the anomalous velocity of the anyon in an electric field.

In a uniform electric field $\bm E$, the Hamiltonian projected onto an isolated band is, up to a constant, $H=E(\bm k)-e^*\bm E\cdot\bm{\mathcal R}$, since the field couples to the dipole moment of the electron density, which in the band is $n_{\rm ex}\bm{\mathcal R}$ [Eq.~\eqref{eq:position-def}].
With $[\mathcal R_i,k_j]=\ii\delta_{ij}$ and Eq.~\eqref{eq:position}, the center $(\bm r_c,\bm k_c)$ of a wave packet narrow in momentum obeys the semiclassical equations, which generalize those of electrons~\cite{Xiao2010} to anyons,
\begin{equation}
 \hbar\dot{\bm k}_c=e^*\bm E,\qquad
 \dot{\bm r}_c=\frac1\hbar\nabla_{\bm k}E(\bm k_c)
 -\dot{\bm k}_c\times\hat{\bm z}\,\Omega(\bm k_c).
 \label{eq:eom}
\end{equation}
Without modulation these give the Hall drift $\dot{\bm r}_c=(3e^*\ell_B^2/\hbar)\,\hat{\bm z}\times\bm E$, and a weak modulation with one flux quantum per cell changes it little [Eq.~\eqref{eq:single-curvature-first}].
We show below that an equally weak modulation with a larger period changes $\Omega(\bm k)$ and the motion qualitatively.

\emph{Splitting into Chern bands of opposite sign.}---We set $\kappa_x=\kappa_y=\kappa$ and double the unit cell by adding two harmonics,
\begin{equation}
 \frac{\delta B(\bm r)}{-B_0}=\kappa\sum_{j=1,2}\cos(\bm b_j\cdot\bm r)
 +\tilde\kappa\sum_{j=1,2}\cos(\tilde{\bm b}_j\cdot\bm r),
 \label{eq:field}
\end{equation}
with $\tilde{\bm b}_1=(\pi/a)(1,1)$ and $\tilde{\bm b}_2=(\pi/a)(-1,1)$.
The primitive translations are now $(a,a)$ and $(-a,a)$, and the unit cell encloses two flux quanta.
The Brillouin zone is halved, and $\tilde{\bm b}_1$ and $\tilde{\bm b}_2$ are now reciprocal vectors, so that $\Gamma\ket{\Psi_{\bm k_1}}$ and $\Gamma\ket{\Psi_{\bm k_2}}$, with $\bm k_1=\bm k$ and $\bm k_2=\bm k+\tilde{\bm b}_1$, have the same crystal momentum under the translations of the doubled cell.
These two states are not orthogonal, and their amplitudes $\bm v=(v_1,v_2)^{\mathsf T}$
in an eigenstate and the two bands $E_\pm(\bm k)$ follow from the generalized eigenvalue problem $H\bm v=ES\bm v$, with ($\alpha,\beta=1,2$)
\begin{equation}
 H_{\alpha\beta}=\bra{\Psi_{\bm k_\alpha}}\Gamma V\Gamma
 \ket{\Psi_{\bm k_\beta}},\;\;
 S_{\alpha\beta}=\bra{\Psi_{\bm k_\alpha}}\Gamma^2
 \ket{\Psi_{\bm k_\beta}}.
 \label{eq:gevp}
\end{equation}
The Hermitian $2\times2$ matrix $H_{\rm eff}=S^{-1/2}HS^{-1/2}$ has the same eigenvalues, with the eigenvectors $S^{1/2}\bm v$.

At weak modulation $H_{\rm eff}$ takes a closed form.
Expanding $\Gamma^2=1-\sum_iw(\bm r_i)+O(w^2)$ in Eq.~\eqref{eq:gevp}, we obtain $H_{\rm eff}=E_0+H^{(1)}+O(w^2)$, with~\cite{SM}
\begin{equation}
 \begin{aligned}
 H^{(1)}_{\alpha\beta}
 &=-\bra{\Psi_{\bm k_\alpha}}(V-E_0)\sum_iw(\bm r_i)
 \ket{\Psi_{\bm k_\beta}}\\
 &=\sum_j\bra{\Psi_{\bm k_\alpha}}
 \frac{\kappa\chi_1}{2\pi}U_{\bm d_j}
 +\frac{\tilde\kappa\tilde\chi}{\pi}U_{\tilde{\bm d}_j}+{\rm H.c.}
 \ket{\Psi_{\bm k_\beta}}.
 \end{aligned}
 \label{eq:heff}
\end{equation}
The second line is for the field of Eq.~\eqref{eq:field}, for which $w=-(\kappa/\pi)\sum_j\cos(\bm b_j\cdot\bm r) -(2\tilde\kappa/\pi)\sum_j\cos(\tilde{\bm b}_j\cdot\bm r)$, and follows from the magnetic translation symmetry of the uniform states.
Here $j=1,2$, $U_{\bm d}$ is the magnetic translation by $\bm d$, with $\bm d_j=3\ell_B^2\,\hat{\bm z}\times\bm b_j$ and $\tilde{\bm d}_j=3\ell_B^2\,\hat{\bm z}\times\tilde{\bm b}_j$, and $\tilde\chi=\bra{\Psi_{\bm 0}}U^\dagger_{\tilde{\bm d}_1}(V-E_0) \rho_{\tilde{\bm b}_1}\ket{\Psi_{\bm 0}}$ is the counterpart of $\chi_1$ for the new harmonics.
The translations $U_{\bm d_j}$ are diagonal in the two states, and $U_{\tilde{\bm d}_j}$ exchange them.
With $\theta_\pm=(3a/2)(k_x\pm k_y)$ and in the basis $e^{\ii\theta_-/2}\ket{\Psi_{\bm k}}$, $e^{-\ii\theta_-/2}U_{\tilde{\bm d}_1}\ket{\Psi_{\bm k}}$, the result is~\cite{SM} $H_{\rm eff}(\bm k)=E_0+\bm h(\bm k)\cdot\bm\sigma$, with
\begin{equation}
 \bm h=\frac2\pi\big(\tilde\kappa\tilde\chi\sin\theta_+,\,
 \tilde\kappa\tilde\chi\sin\theta_-,\,
 \kappa\chi_1\cos\theta_+\cos\theta_-\big).
 \label{eq:two-level}
\end{equation}
The bands $E_\pm=E_0\pm|\bm h|$ have the direct gap $(4/\pi)\min(|\kappa\chi_1|,|\tilde\kappa\tilde\chi|)$.

When the gap is open, the curvature of each band is defined as for the single band, by the form factor of Eq.~\eqref{eq:form-factor} between the band eigenstates $\ket{\Psi^\pm_{\bm k}}\propto\Gamma\sum_\alpha v^\pm_\alpha(\bm k) \ket{\Psi_{\bm k_\alpha}}$.
To leading order $\Gamma\to1$, and $\rho_{\bm q}$ does not mix the two uniform components.
In the basis of Eq.~\eqref{eq:two-level} the form factor then factorizes into that of the uniform band, $\Lambda^{(0)}_{\bm q}(\bm k)$, times the overlap $\bm v^{\pm\dagger}(\bm k+\bm q)\,\bm v^\pm(\bm k)$ of the eigenvectors $\bm v^\pm$ of $\bm h\cdot\bm\sigma$.
Their phases add, and~\cite{SM}
\begin{equation}
 \Omega^{(0)}_\mp=3\ell_B^2\pm\frac12\hat{\bm h}\cdot
 (\partial_{k_x}\hat{\bm h}\times\partial_{k_y}\hat{\bm h}),
 \label{eq:curvature}
\end{equation}
with the upper sign for the lower band, $\hat{\bm h}=\bm h/|\bm h|$, and corrections proportional to $\kappa$ and $\tilde\kappa$~\cite{SM}.
Over the halved Brillouin zone the two terms of Eq.~\eqref{eq:curvature} contribute $3/2$ and $\mp(9/2)\operatorname{sgn}(\kappa\chi_1)$ to the Chern number, so that
\begin{equation}
 C_\mp=\frac32\mp\frac92\operatorname{sgn}(\kappa\chi_1),
 \qquad C_-+C_+=3.
 \label{eq:chern}
\end{equation}
For $\kappa\chi_1<0$ the lower and the upper band have $C=6$ and $-3$.
Since $\hat{\bm h}$ depends only on the ratio of the two amplitudes, an arbitrarily weak modulation splits the band into two bands with Chern numbers of opposite sign.
This pattern is fixed by the fractional charge: we show in the SM that the magnetic translation algebra of the quasihole states requires $C\equiv6\pmod9$ for each band whenever a gap opens, a Diophantine constraint~\cite{Thouless1982,Dana1985} for a charge $e/3$ that sees $2/3$ of a flux quantum per cell, so that $(6,-3)$ is the pair of smallest magnitude; for an electron the constraint is empty, and its Landau level with $C=-1$ splits into bands with $C=-1$ and $0$.

\begin{figure}[t]
\centering
\includegraphics[width=\columnwidth]{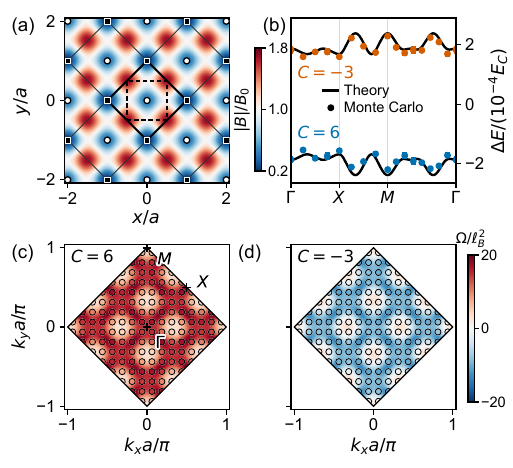}
\caption{Split quasihole bands at $(\kappa,\tilde\kappa)=(-0.35,0.05)$ and $d=\ell_B$, with Monte Carlo data at $(N_\phi,N_e)=(256,85)$.
(a) Field magnitude: solid diamonds (dashed square) enclose cells of area $2a^2$ ($a^2$); open circles (filled squares) mark minima of $0.4B_0$ ($0.2B_0$).
(b) Energies $\Delta E=E-\overline E$, with $\overline E$ the mean over both bands and momenta; curves show $E_0\pm|\bm h|$.
(c),(d) Lower- and upper-band density-link curvature; backgrounds show Eq.~\eqref{eq:curvature}.
Points are Monte Carlo data; error bars denote one bootstrap standard deviation.
Theory uses uniform-state coefficients.}
\label{fig:static}
\end{figure}

We test these results by a Monte Carlo evaluation of Eq.~\eqref{eq:gevp} and of the density links~\cite{Fukui2005,SM}, for the screened Coulomb interaction with $d=\ell_B$ and $(\kappa,\tilde\kappa)=(-0.35,0.05)$, for which the field varies between $0.2B_0$ and $1.7B_0$ [Fig.~\ref{fig:static}(a)].
The two bands have the Chern numbers $(C_-,C_+)=(6,-3)$, as in Eq.~\eqref{eq:chern} with $\kappa\chi_1<0$.
The leading-order bands and continuous curvature, evaluated with the uniform-state coefficients, agree well with the Monte Carlo results [Figs.~\ref{fig:static}(b)--\ref{fig:static}(d)].

\begin{figure}[t]
\centering
\includegraphics[width=\columnwidth]{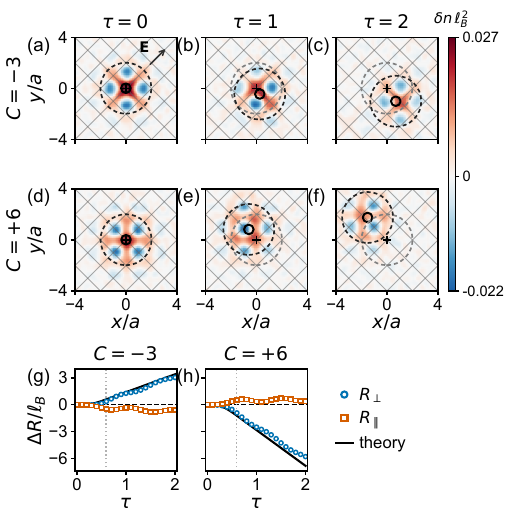}
\caption{Quasihole drift at $(N_\phi,N_e)=(256,85)$, with other parameters as in Fig.~\ref{fig:static}.
(a)--(f) Density deficit for packets initially in the upper (top) and lower (bottom) band.
Dashed circles mark the evolving (black) and initial (gray) charge centers.
(g),(h) Transverse and parallel displacements relative to the electric field; curves show Eq.~\eqref{eq:mean-drift}.}
\label{fig:dynamics}
\end{figure}

\emph{Reversal of the drift.}---We study the dynamics of a quasihole wave packet under an electric field $\bm E$.
We neglect the group velocity of the nearly flat bands [Fig.~\ref{fig:static}(b)] and approximate the Berry curvature by its Brillouin-zone average $2C\ell_B^2$.
Equation~\eqref{eq:eom} then gives
\begin{equation}
 \dot{\bm{\mathcal R}}=2C\,\frac{e^*\ell_B^2}{\hbar}\,
 \hat{\bm z}\times\bm E.
 \label{eq:mean-drift}
\end{equation}
A wave packet in the $C=6$ band drifts twice as fast as one in the $C=-3$ band, in the opposite direction.

We test this prediction with the microscopic wave functions, at the parameters of Fig.~\ref{fig:static}~\cite{SM}.
The initial state is a quasihole localized at the origin and projected onto one band,
\begin{equation}
 \ket{\zeta_\pm}\propto P_\pm\,\Gamma\ket{\psi^\sigma_{\eta=0}},
 \qquad
 P_\pm=\sum_{\bm k}\ket{\Psi^\pm_{\bm k}}\bra{\Psi^\pm_{\bm k}},
 \label{eq:initial-packet}
\end{equation}
with $\psi^\sigma_\eta$ of Eq.~\eqref{eq:torus-qh} and the normalized band states $\ket{\Psi^\pm_{\bm k}}$.
The electric field is produced by the potential energy of the electrons
\begin{equation}
 \hat V_{\mathcal E}(t)=\frac{e\,\mathcal E(t)}{\sqrt2\,q}
 \sum_i\sin[q(x_i+y_i)],\qquad q=\frac{2\pi}{L},
 \label{eq:electric-drive}
\end{equation}
where $L$ is the side of the torus.
A uniform field has a linear potential, which is not periodic on the torus.
The wave vector $q$ makes $\hat V_{\mathcal E}$ periodic, and for $q|x_i+y_i|\ll1$ it reduces to the potential energy $e\mathcal E(t)\sum_i(x_i+y_i)/\sqrt2$ of the electrons in a uniform field $\mathcal E(t)$ along $(1,1)/\sqrt2$.
The field is switched on as $\mathcal E(t)=\tfrac12\mathcal E[1-\cos(\pi\tau/\tau_r)]$ for $\tau<\tau_r$ and stays at $\mathcal E$ afterwards, with the time $\tau=e\mathcal E\ell_Bt/\hbar$~\cite{SM}.
Per unit of $\tau$, a quasihole in the uniform field moves by $\ell_B$, and by Eq.~\eqref{eq:mean-drift} the packets of the upper and the lower band move by $2\ell_B$ and $4\ell_B$, in opposite directions.

We evolve the state in the space of the two bands and follow the center of the density deficit.
The packets move in opposite directions, that of the upper band opposite to $\hat{\bm z}\times\bm E$ [Figs.~\ref{fig:dynamics}(a)--\ref{fig:dynamics}(f)], and their displacements follow Eq.~\eqref{eq:mean-drift} [Figs.~\ref{fig:dynamics}(g) and \ref{fig:dynamics}(h)]: at $\tau=2$ they are $3.05(5)\ell_B$ and $5.79(8)\ell_B$, against the predicted $3.41\ell_B$ and $6.83\ell_B$, about one and two lattice constants $\sqrt2a$ of the doubled cell.
We attribute the difference, and the slight motion along the field, to the two approximations in Eq.~\eqref{eq:mean-drift}.
The magnetic field has the same sign everywhere, between $0.2B_0$ and $1.7B_0$, so the reversal reflects the geometry of the band.

\emph{Summary and Discussion.}---For a Laughlin quasihole in a periodically modulated magnetic field, we have derived the guiding-center commutator, Eq.~\eqref{eq:position}, and the wave-packet equations of motion, Eq.~\eqref{eq:eom}, from microscopic wave functions.
A modulation with two flux quanta per unit cell splits the parent band with Chern number $3$ into bands with Chern numbers $6$ and $-3$.
We demonstrate that a wave packet in the $C=-3$ band drifts against the $\bm E\times\bm B$ direction even though the magnetic field has the same sign everywhere.
Monte Carlo calculations with the microscopic wave functions allow us to reach large systems ($N_e=85$).
Exact diagonalization in smaller systems gives consistent results~\cite{SM}.
Perturbation theory in the modulation strength explains the band splitting and provides a quantitative description of the dispersions and Berry curvatures.

The emergence of opposite-sign Chern bands is not specific to the square lattice.
A triangular modulation produces the same pair of Chern numbers, with $C=-3$ in the lower band, and a suitably chosen weak periodic scalar potential realizes the same splitting at first order~\cite{SM}.

For doped quasiholes, the lower band can have $C=-3$, for example for $\kappa\chi_1>0$ in Eq.~\eqref{eq:chern}, and filling it would enhance rather than reduce the Hall response of the parent state.
It would be interesting to study whether such band geometry leads to correlated phases whose Hall conductance is not fixed by the filling~\cite{Kol1993,Sohal2018,Sarkar2025}, and how it affects anyon pairing~\cite{Li2026a,Pichler2026,Fan2026WeakPairing}.

The composite-fermion construction~\cite{Jain1989} of quasiparticle and quasihole bands in Jain states with modulated magnetic field~\cite{Pu2026} provides a microscopic basis for extending this framework beyond Laughlin quasiholes.
The same magnetic translation structure suggests that enlarging the modulation cell can similarly split these bands into subbands with Chern numbers of opposite sign.

Periodic strain~\cite{Mao2020,Wan2023,Gao2023} can produce flat bands equivalent to a Landau level with two flux quanta per unit cell~\cite{Sarkar2025}, and patterned electrostatic superlattices~\cite{Forsythe2018,BarconsRuiz2022,Ghorashi2023,Wang2024a,Sun2024,Tan2024a,Shi2025a,Wang2026a} offer a tunable alternative through the scalar potential noted above.
Tunneling probes have been proposed to measure the anyon dispersion~\cite{Wang2026}, localized fractionally charged states have been imaged~\cite{Martin2004}, and anyon tunneling at quantum Hall edges has been probed in the time domain~\cite{Ruelle2025}.
These developments motivate experimental studies of the bulk quasihole dynamics predicted here.

\emph{Acknowledgment.}--- We thank Biao Lian, Miguel {\fontencoding{T1}\selectfont Gon\c{c}alves}, and Takamori Park for helpful discussions.
Z.X. thanks Zhi-Da Song for helpful discussions.
The computations reported in this paper were performed using Princeton Research Computing resources.
Z.X. is supported by the Princeton Quantum Initiative Fellowship.
S.R. is supported by a Simons
Investigator Grant from the Simons Foundation 
(Award No.\ 566116). This work is supported by the Gordon and Betty
Moore Foundation EPiQS initiative, Grant GBMF8685.01.
The authors used OpenAI Codex (GPT-6 Astra) and Claude Code (Opus 5) for code development and debugging, checks of analytical steps, and manuscript revision.
The authors critically reviewed and independently verified all outputs through analytical and numerical checks, and retain full responsibility for this work.

\bibliography{references,main_refs}

\clearpage \section*{End Matter} \appendix* \setcounter{equation}{0} \renewcommand{\theequation}{A\arabic{equation}}

\begin{figure}[b]
\centering
\includegraphics[width=\columnwidth]{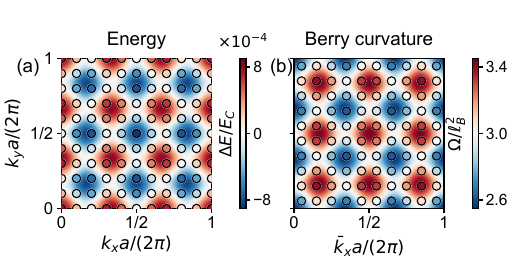}
\caption{Single quasihole band at $\kappa_x=\kappa_y=0.35$, $d=5\ell_B$, and $(N_\phi,N_e)=(100,33)$.
(a) $\Delta E=E-\overline E$, with $\overline E$ the momentum average and $E_C=e^2/(\epsilon\ell_B)$.
(b) Density-link curvature.
Circles: Monte Carlo; backgrounds: Eqs.~\eqref{eq:single-dispersion} and \eqref{eq:single-curvature-first}, with uniform-state coefficients.}
\label{fig:single-band}
\end{figure}

\phantomsection\label{sec:endmatter-single}
\emph{Appendix A: Single band with one flux quantum per cell.}---Since $\Gamma$ and $V$ are both diagonal in the electron positions, $\Gamma V\Gamma=Ve^{-W}$ and $\Gamma^2=e^{-W}$, with
$W=\sum_iw(\bm r_i)=\sum_{\bm G}w_{\bm G}\,\rho_{\bm G}$, where $w_{\bm G}$ are the Fourier components of $w$ at the reciprocal vectors $\bm G$.
The energy [Eq.~\eqref{eq:single-energy}] and the normalization are therefore expectation values in the uniform state,
\begin{equation}
 E(\bm k)=\frac{\langle Ve^{-W}\rangle_{\bm k}}{\langle e^{-W}\rangle_{\bm k}},
 \qquad
 \mathcal N(\bm k)=\langle e^{-W}\rangle_{\bm k},
 \label{eq:em-single-ratio}
\end{equation}
with $\langle A\rangle_{\bm k}=\bra{\Psi_{\bm k}}A\ket{\Psi_{\bm k}}$ in the normalized anyon Bloch state of the uniform field [Eq.~\eqref{eq:bloch}].
Since the band is flat without modulation, $\langle V\rangle_{\bm k}=E_0$ at every $\bm k$, and to first order in $w$
\begin{equation}
 \begin{aligned}
 E(\bm k)-E_0&=-\big\langle(V-E_0)W\big\rangle_{\bm k},\\
 \ln\mathcal N(\bm k)&=-\langle W\rangle_{\bm k}.
 \end{aligned}
 \label{eq:em-single-first}
\end{equation}
In $\mathcal H_{\rm qh}$, which holds one state $\ket{\Psi_{\bm k}}$ per momentum, $\rho_{\bm q}$ acts as the magnetic translation $U_{\bm d(\bm q)}$, $\bm d(\bm q)=3\ell_B^2\hat{\bm z}\times\bm q$, times a $\bm k$-independent factor~\cite{SM}.
Since $V$ commutes with the magnetic translations, the same holds for $(V-E_0)\rho_{\bm q}$.
For a reciprocal vector $\bm G=m_1\bm b_1+m_2\bm b_2$, $\bm d(\bm G)=3(m_1\bm a_2-m_2\bm a_1)$ is a lattice vector; up to a constant phase, $U_{\bm d(\bm G)}$ is $U_{\bm a_2}^{3m_1}U_{\bm a_1}^{-3m_2}$ and multiplies $\ket{\Psi_{\bm k}}$ by $e^{-\ii\bm k\cdot\bm d(\bm G)}$.
Hence the momentum dependence of both expectation values is a phase,
\begin{equation}
 \begin{aligned}
 \langle\rho_{\bm G}\rangle_{\bm k}&=\langle\rho_{\bm G}\rangle_{\bm 0}\,e^{-\ii\bm k\cdot\bm d(\bm G)},\\
 \big\langle(V-E_0)\rho_{\bm G}\big\rangle_{\bm k}&=\chi(\bm G)\,e^{-\ii\bm k\cdot\bm d(\bm G)},
 \end{aligned}
 \label{eq:em-single-phase}
\end{equation}
with $\chi(\bm G)=\langle(V-E_0)\rho_{\bm G}\rangle_{\bm 0}$; both $\langle\rho_{\bm G}\rangle_{\bm 0}$ and $\chi(\bm G)$ are real and even in $\bm G$.

Inserting these phases into the first-order expressions [Eq.~\eqref{eq:em-single-first}], and $\ln\mathcal N$ into the curvature formula [Eq.~\eqref{eq:single-curvature}], gives
\begin{equation}
 \begin{aligned}
 E(\bm k)-E_0&=-\sum_{\bm G}w_{\bm G}\,\chi(\bm G)\,e^{-\ii\bm k\cdot\bm d(\bm G)},\\
 \Omega(\bm k)-3\ell_B^2&=-\frac12\sum_{\bm G}w_{\bm G}\,\langle\rho_{\bm G}\rangle_{\bm 0}\,|\bm d(\bm G)|^2\,e^{-\ii\bm k\cdot\bm d(\bm G)}.
 \end{aligned}
 \label{eq:em-single-general}
\end{equation}
For the first-harmonic modulation of the main text, $w=-[\kappa_x\cos(\bm b_1\cdot\bm r)+\kappa_y\cos(\bm b_2\cdot\bm r)]/\pi$, so that $w_{\pm\bm b_1}=-\kappa_x/2\pi$ and $w_{\pm\bm b_2}=-\kappa_y/2\pi$, with $\bm d(\bm b_1)=3a\hat{\bm y}$, $\bm d(\bm b_2)=-3a\hat{\bm x}$, and $|\bm d(\bm b_j)|^2=18\pi\ell_B^2$.
With $\chi(\pm\bm b_j)=\chi_1$ and $3\langle\rho_{\pm\bm b_j}\rangle_{\bm 0}=\xi_1$ by the fourfold symmetry of the uniform state, these sums give the dispersion and the curvature of the main text [Eqs.~\eqref{eq:single-dispersion} and \eqref{eq:single-curvature-first}].
Both repeat nine times in the Brillouin zone because $|\bm d(\bm b_j)|=3a$, a generic feature at filling $1/3$~\cite{Yan2026,Iyer2026}.
A Monte Carlo evaluation of the energy [Eq.~\eqref{eq:em-single-ratio}] and of the density-link curvature [Eq.~\eqref{eq:plaquette}] agrees with both, with $\chi_1$ and $\xi_1$ computed in the uniform state and no fitted parameters [Fig.~\ref{fig:single-band}], although the modulation is not weak: the field varies between $0.3B_0$ and $1.7B_0$.

\phantomsection\label{sec:endmatter-dynamics}
\emph{Appendix B: Wave-packet dynamics of a quasihole.}---We work in the bulk thermodynamic limit and in a smooth momentum patch of one isolated band.
Write $\ket\zeta=\int_{\bm k}\zeta(\bm k) \ket{\Psi^\Gamma_{\bm k}}$, with $\int_{\bm k}|\zeta|^2=1$, where $\int_{\bm k}$ is the continuum limit of the Bloch momentum sum.
The density deficit relative to the ground state in the same magnetic field is $\delta n_\zeta(\bm r)=\bra\zeta\hat n(\bm r)\ket\zeta-n_{\rm bg}(\bm r)$.
We assume a localized packet with finite absolute second density moment and fixed excess electron number.
Its charge and physical center are
\begin{equation}
 \begin{aligned}
 n_{\rm ex}&=\int\dd^2r\,\delta n_\zeta=-\frac13,
 \qquad e^*=-e\,n_{\rm ex}=\frac e3,\\
 \bm r_c&=\langle\bm{\mathcal R}\rangle_\zeta
 =\frac1{n_{\rm ex}}\int\dd^2r\,\bm r\,\delta n_\zeta(\bm r).
 \end{aligned}
 \label{eq:em-centroid}
\end{equation}

For a nonzero $\bm q$ that is not a reciprocal vector of the modulation, with $|\bm q|\ell_\zeta\ll1$, where $\ell_\zeta$ is the size of the packet, we evaluate $\langle\rho_{\bm q}\rangle_\zeta$ in two ways.
In real space,
\begin{equation}
 \begin{aligned}
 \langle\rho_{\bm q}\rangle_\zeta
 &=\int\dd^2r\,e^{\ii\bm q\cdot\bm r}
 \big[n_{\rm bg}(\bm r)+\delta n_\zeta(\bm r)\big]\\
 &=\int\dd^2r\,(1+\ii\bm q\cdot\bm r)\,\delta n_\zeta(\bm r)+O(q^2)\\
 &=n_{\rm ex}\big[1+\ii\bm q\cdot\bm r_c\big]+O(q^2),
 \end{aligned}
 \label{eq:em-density-split}
\end{equation} 
where $n_{\rm bg}(\bm r)$ has no Fourier component at a small but nonzero $\bm q$ because of its periodicity.
In momentum space, $\rho_{\bm q}$ changes the momentum by $\bm q$, so that
\begin{equation}
 \langle\rho_{\bm q}\rangle_\zeta
 =\int_{\bm k}\zeta^*(\bm k+\bm q)\,
 \Lambda_{\bm q}(\bm k)\,\zeta(\bm k),
 \label{eq:em-bloch}
\end{equation}
with the form factor $\Lambda_{\bm q}(\bm k)$ [Eq.~\eqref{eq:form-factor}].
We compare the two expressions order by order in $\bm q$.
At zeroth order, the comparison implies that $\Lambda_{\bm q}(\bm k)\to n_{\rm ex}$ as $\bm q\to0$ ($\bm q \neq 0$).
At first order, we write $\Lambda_{\bm q}(\bm k)=n_{\rm ex} [1+\bm q\cdot\bm\lambda(\bm k)+O(q^2)]$.
Since $\rho_{\bm q}^\dagger=\rho_{-\bm q}$, the form factor obeys $\Lambda_{-\bm q}(\bm k+\bm q)=\Lambda_{\bm q}(\bm k)^*$, which at this order reads $-\bm\lambda=\bm\lambda^*$.
Hence $\bm\lambda=\ii\bm{\mathcal A}$ with a real $\bm{\mathcal A}$,
\begin{equation}
 \Lambda_{\bm q}(\bm k)=n_{\rm ex}
 \big[1+\ii\bm q\cdot\bm{\mathcal A}(\bm k)+O(q^2)\big].
 \label{eq:em-link}
\end{equation}
We insert Eq.~\eqref{eq:em-link} and $\zeta^*(\bm k+\bm q)=\zeta^*(\bm k) +\bm q\cdot\nabla_{\bm k}\zeta^*(\bm k)+O(q^2)$ into Eq.~\eqref{eq:em-bloch} and integrate the gradient by parts, $\int_{\bm k}(\bm q\cdot\nabla_{\bm k}\zeta^*)\,\zeta =-\int_{\bm k}\zeta^*\,\bm q\cdot\nabla_{\bm k}\zeta$, where $\zeta$ vanishes at the boundary of the patch:
\begin{equation}
 \langle\rho_{\bm q}\rangle_\zeta=n_{\rm ex}\Big[1+\ii\bm q\cdot\!
 \int_{\bm k}\zeta^*\big(\ii\nabla_{\bm k}+\bm{\mathcal A}\big)\zeta
 \Big]+O(q^2).
 \label{eq:em-moment}
\end{equation}
Comparison with Eq.~\eqref{eq:em-density-split} gives $\langle\bm{\mathcal R}\rangle_\zeta=\int_{\bm k}\zeta^* (\ii\nabla_{\bm k}+\bm{\mathcal A})\zeta$ for every packet in the patch.
Both sides are expectation values of Hermitian operators, which are therefore equal,
\begin{equation}
 \bm{\mathcal R}=\ii\nabla_{\bm k}+\bm{\mathcal A}(\bm k).
 \label{eq:em-position-bloch}
\end{equation}

The curl of $\bm{\mathcal A}$ is the curvature of Eq.~\eqref{eq:plaquette}.
By Eq.~\eqref{eq:em-link} and $\Lambda_{-\bm q}(\bm k+\bm q)=\Lambda_{\bm q}(\bm k)^*$, the phase of the four form factors around the plaquette is, with $q_x=|\bm q_x|$ and $q_y=|\bm q_y|$,
\begin{align}
 &q_x\mathcal A_x(\bm k)+q_y\mathcal A_y(\bm k+\bm q_x)
 -q_x\mathcal A_x(\bm k+\bm q_y)-q_y\mathcal A_y(\bm k)\nonumber\\
 &\quad=q_xq_y\,
 (\partial_{k_x}\mathcal A_y-\partial_{k_y}\mathcal A_x)
 \label{eq:em-plaquette}
\end{align}
to leading order in the mesh spacing; the four factors of $n_{\rm ex}$ have no net phase.
Division by the area $q_xq_y$ gives $\Omega=\partial_{k_x}\mathcal A_y-\partial_{k_y}\mathcal A_x$, and Eq.~\eqref{eq:em-position-bloch} then gives the commutators
\begin{equation}
 \begin{aligned}
 \relax[\mathcal R_x,\mathcal R_y]
 &=[\ii\partial_{k_x}+\mathcal A_x,\,\ii\partial_{k_y}+\mathcal A_y]\\
 &=\ii(\partial_{k_x}\mathcal A_y-\partial_{k_y}\mathcal A_x)
 =\ii\Omega(\bm k),\\
 \relax[\mathcal R_a,k_b]&=\ii\delta_{ab}.
 \end{aligned}
 \label{eq:em-commutators}
\end{equation}
Under $\ket{\Psi^\Gamma_{\bm k}}\to e^{\ii\theta(\bm k)} \ket{\Psi^\Gamma_{\bm k}}$, the connection changes by $-\nabla_{\bm k}\theta$ and $\zeta\to e^{-\ii\theta}\zeta$, so the position is gauge covariant.

A uniform electric field, $\phi(\bm r)=\phi_0-\bm E\cdot\bm r$, couples to the electrons through
\begin{equation}
 \begin{aligned}
 H_{\rm ext}&=-e\!\int\!\dd^2r\,\phi(\bm r)\,\hat n(\bm r)\\
 &=-e\phi_0N_e+e\bm E\cdot\!\int\!\dd^2r\,\bm r\,\hat n(\bm r).
 \end{aligned}
 \label{eq:em-external}
\end{equation}
By Eq.~\eqref{eq:position-def}, the projection of the last integral onto the band is $n_{\rm ex}\bm{\mathcal R}$ plus the constant contribution of the background.
With $e^*=-en_{\rm ex}$ and up to a constant, the Hamiltonian in the band is therefore
\begin{equation}
 H_{\rm band}=E(\bm k)-e^*\bm E\cdot\bm{\mathcal R}.
 \label{eq:em-hamiltonian}
\end{equation}
The Heisenberg equations $\dot O=(\ii/\hbar)[H_{\rm band},O]$, with Eq.~\eqref{eq:em-commutators} and $[E(\bm k),\mathcal R_a]=-\ii\partial_{k_a}E$, are
\begin{equation}
 \begin{aligned}
 \hbar\dot{\bm k}&=e^*\bm E,\\
 \hbar\dot{\bm{\mathcal R}}&=\nabla_{\bm k}E(\bm k)
 -e^*\bm E\times\hat{\bm z}\,\Omega(\bm k).
 \end{aligned}
 \label{eq:em-heisenberg}
\end{equation}

\end{document}


\title{\textbf{\large Supplemental Material for ``Anyon Chern bands of opposite sign and reversed drift''}}

\author{Zhenyu Xiao}
\email{zyxiao@princeton.edu}
\affiliation{Princeton Quantum Initiative, Princeton University, Princeton, New Jersey 08544, USA}
\affiliation{Princeton Center for Theoretical Science, Princeton University, Princeton, New Jersey 08544, USA}

\author{Minxuan Wang}
\affiliation{Department of Physics, Princeton University, Princeton, New Jersey 08544, USA}

\author{Shinsei Ryu}
\affiliation{Department of Physics, Princeton University, Princeton, New Jersey 08544, USA}

\maketitle

\setcounter{secnumdepth}{3} \renewcommand{\thesection}{S\Roman{section}} \renewcommand{\thesubsection}{\thesection.\Alph{subsection}} \renewcommand{\thesubsubsection}{\thesubsection.\arabic{subsubsection}} \makeatletter \renewcommand{\p@subsection}{} \renewcommand{\p@subsubsection}{} \makeatother \renewcommand{\theequation}{S\arabic{equation}} \renewcommand{\thefigure}{S\arabic{figure}} \renewcommand{\thetable}{S\arabic{table}} \setcounter{section}{0} \setcounter{subsection}{0} \setcounter{equation}{0} \setcounter{figure}{0} \setcounter{table}{0}

\setcounter{tocdepth}{2} \tableofcontents 
\section{Quasihole states on the torus} \label{sec:anyon-bloch} \label{sec:ac-qh}

\subsection{Laughlin quasihole and boundary conditions} \label{sec:torus-setup} \label{sec:torus-localized}

Electrons of charge $-e$ move in the field $-B_0\hat{\bm z}$, with $\ell_B=\sqrt{\hbar/eB_0}$, on a torus $L_x\times L_y$ divided into $N_1\times N_2$ cells $a_1\times a_2$ that each enclose one flux quantum, $a_1a_2=2\pi\ell_B^2$.
The main text uses $a_1=a_2=a$.
With $L_x=N_1a_1$, $L_y=N_2a_2$, and $N_\phi=N_1N_2=3N_e+1$, the boundary conditions in the symmetric gauge $\bm A=(B_0y/2,-B_0x/2)$ are
\begin{equation}
 \Phi(\ldots,z_i+L_x,\ldots)
 =e^{\ii L_xy_i/(2\ell_B^2)}\Phi(\{z_i\}),\qquad
 \Phi(\ldots,z_i+\ii L_y,\ldots)
 =e^{-\ii L_yx_i/(2\ell_B^2)}\Phi(\{z_i\}),
 \label{eq:torus-electron-bc}
\end{equation}
with $z_i=x_i+\ii y_i$.
The Laughlin state with one quasihole at $\eta=X+\ii Y$ is~\cite{Haldane1985a,Greiter2016}
\begin{equation}
 \psi^\sigma_\eta=e^{(\eta^2-|\eta|^2)/(12\ell_B^2)}\,
 e^{\sum_i(z_i^2-|z_i|^2)/(4\ell_B^2)}\,F^\sigma\big(Z+\tfrac\eta3\big)
 \prod_i\vartheta_1\!\Big(\frac{z_i-\eta}{L_x}\,\Big|\,\tau\Big)
 \prod_{i<j}\vartheta_1\!\Big(\frac{z_i-z_j}{L_x}\,\Big|\,\tau\Big)^3,
 \qquad \tau=\frac{\ii L_y}{L_x},
 \label{eq:torus-qh-sm}
\end{equation}
with $Z=\sum_iz_i$ and the center-of-mass factor
\begin{equation}
 F^\sigma(W)=e^{-4\pi\ii N_e\sigma/3}
 \vartheta\!\left[\begin{matrix}\sigma/3+N_e/2\\N_e/2\end{matrix}\right]
 \left(\frac{3W}{L_x}\,\middle|\,3\tau\right),\qquad
 \sigma=0,1,2,
 \label{eq:torus-com}
\end{equation}
where
\begin{equation}
 \vartheta\!\left[\begin{matrix}\alpha\\\beta\end{matrix}\right](w\mid\tau)
 =\sum_{n\in\mathbb Z}
 e^{\pi\ii\tau(n+\alpha)^2+2\pi\ii(n+\alpha)(w+\beta)}.
 \label{eq:torus-theta}
\end{equation}
The quasiperiodicity of the theta functions gives, with $\ell_h=\sqrt3\ell_B$,
\begin{equation}
 \ket{\psi^\sigma_{\eta+L_x}}
 =e^{2\pi\ii\sigma/3}e^{\ii L_xY/(2\ell_h^2)}\ket{\psi^\sigma_\eta},
 \qquad
 \ket{\psi^\sigma_{\eta+\ii L_y}}
 =e^{-\ii L_yX/(2\ell_h^2)}\ket{\psi^{\sigma+1}_\eta},
 \label{eq:torus-sector-bc}
\end{equation}
with $\sigma+3\equiv\sigma$.
The three sectors thus belong to one family as the quasihole traverses the torus.
For every $\eta$ and $\sigma$, $\psi^\sigma_\eta$ is a zero mode of the Trugman--Kivelson (TK) interaction~\cite{Trugman1985}, and these states span the single-quasihole space $\mathcal H_{\rm qh}$, of dimension $N_\phi$~\cite{Regnault2011}.

\subsection{Anyon Bloch states} \label{sec:torus-bloch}

The magnetic translation $U_{\bm c}$ moves every electron by $\bm c$,
$(U_{\bm c}\Phi)(\{z_i\})=e^{\ii\operatorname{Im}(\bar cZ)/(2\ell_B^2)}\,\Phi(\{z_i-c\})$ with $c=c_x+\ii c_y$, and
\begin{equation}
 U_{\bm c}U_{\bm c'}
 =e^{\ii N_e(\bm c\times\bm c')_z/\ell_B^2}
 U_{\bm c'}U_{\bm c}.
 \label{eq:translation-algebra}
\end{equation}
The smallest translations preserving the magnetic boundary conditions [Eq.~\eqref{eq:torus-electron-bc}] are the fine translations
\begin{equation}
 T_x=U_{(L_x/N_\phi)\hat{\bm x}},\qquad
 T_y=U_{(L_y/N_\phi)\hat{\bm y}}.
 \label{eq:fine-translations}
\end{equation}
They preserve $\mathcal H_{\rm qh}$, since the TK interaction is translation invariant, and give the primitive translations $U_{\bm a_1}=T_x^{N_2}$ and $U_{\bm a_2}=T_y^{N_1}$, with $\bm a_1=a_1\hat{\bm x}$ and $\bm a_2=a_2\hat{\bm y}$.
These commute, and their eigenvalues $e^{-\ii\bm k\cdot\bm a_j}$ define the total momentum $\bm k$ on the grid
\begin{equation}
 \bm k=\frac{n_1}{N_1}\bm b_1+\frac{n_2}{N_2}\bm b_2,
 \qquad \bm b_1=\frac{2\pi}{a_1}\hat{\bm x},\quad
 \bm b_2=\frac{2\pi}{a_2}\hat{\bm y},
 \qquad n_j=0,\ldots,N_j-1.
 \label{eq:torus-k-grid}
\end{equation}
Let $\mathcal H_{\bm k}\subset\mathcal H_{\rm qh}$ be the eigenspace with momentum $\bm k$.
To establish that each momentum occurs exactly once, use
\begin{equation}
 U_{\bm a_1}T_y=e^{2\pi\ii N_e/N_1}T_yU_{\bm a_1},\qquad
 U_{\bm a_2}T_x=e^{-2\pi\ii N_e/N_2}T_xU_{\bm a_2}.
 \label{eq:torus-fine-shift}
\end{equation}
Thus $T_y$ sends $\mathcal H_{\bm k}$ to $\mathcal H_{\bm k-(N_e/N_1)\bm b_1}$, and $T_x$ sends it to $\mathcal H_{\bm k+(N_e/N_2)\bm b_2}$.
Since $N_1$ and $N_2$ divide $N_\phi=3N_e+1$, $N_e$ is coprime to both, and these maps connect every momentum.
All $\mathcal H_{\bm k}$ therefore have the same dimension, and with $\dim\mathcal H_{\rm qh}=N_\phi$,
\begin{equation}
 \dim\mathcal H_{\bm k}=1.
 \label{eq:torus-qh-momenta}
\end{equation}
The Bloch sum
\begin{equation}
 \ket{\Psi_{\bm k}}\propto\sum_{r=0}^{N_1-1}\sum_{t=0}^{N_2-1}
 e^{\ii\bm k\cdot(r\bm a_1+t\bm a_2)}\,
 U_{\bm a_1}^rU_{\bm a_2}^t\ket{\psi^\sigma_\eta}
 \label{eq:torus-bloch-sm}
\end{equation}
projects onto $\mathcal H_{\bm k}$, so that it gives the same normalized state up to a phase for any $\eta$ and $\sigma$ with nonzero projection.
Moreover, a subspace invariant under $T_x$ and $T_y$ is invariant under $U_{\bm a_1}$ and $U_{\bm a_2}$, contains some $\mathcal H_{\bm k}$, and hence all of them: the representation of $T_x$ and $T_y$ on $\mathcal H_{\rm qh}$ is irreducible~\cite{Haldane1985b}.

\subsection{Quasiholes in the Aharonov--Casher band} \label{sec:ac-band}

We now modulate the field, $B_z(\bm r)=-B_0+\delta B(\bm r)$, where $\delta B$ is periodic with zero average, so that the total flux $N_\phi$ is unchanged.
For an electron with a $g$ factor of $2$ in the spin-polarized branch, the Pauli term combines with the kinetic energy into the Aharonov--Casher (AC) Hamiltonian~\cite{Aharonov1979,Shi2024},
\begin{equation}
 h_{\rm AC}=\frac{\bm\pi^2+e\hbar B_z}{2m_e}
 =\frac{D^\dagger D}{2m_e},\qquad D=\pi_x+\ii\pi_y,
 \label{eq:ac-hamiltonian}
\end{equation}
with $\bm\pi=-\ii\hbar\nabla+e\bm A$ and $[\pi_x,\pi_y]=-\ii e\hbar B_z$.
Since $h_{\rm AC}\geq0$, its zero modes are the solutions of $D\varphi=0$.
They form the AC band, which is exactly flat for any $\delta B$.
To find them, write the change of the vector potential as $\delta\bm A=-(\hbar/2e)\hat{\bm z}\times\nabla w$ with a periodic function $w$, so that $\nabla^2w=-(2e/\hbar)\delta B$.
Then
\begin{equation}
 D=e^{-w/2}D_0e^{w/2},\qquad
 D(e^{-w/2}\varphi_0)=e^{-w/2}D_0\varphi_0=0,
 \label{eq:ac-zero-mode}
\end{equation}
where $D_0$ is the operator in the uniform field and $\varphi_0$ is a lowest-Landau-level orbital.
Multiplication by the envelope $e^{-w/2}$ thus maps the lowest Landau level one-to-one onto the AC band, which therefore has $N_\phi$ states.
For $\delta B=-B_0\sum_{\bm G\ne0}\kappa_{\bm G}e^{\ii\bm G\cdot\bm r}$,
\begin{equation}
 w(\bm r)=\sum_{\bm G\ne0}w_{\bm G}e^{\ii\bm G\cdot\bm r},\qquad
 w_{\bm G}=-\frac{2\kappa_{\bm G}}{\ell_B^2|\bm G|^2}.
 \label{eq:ac-poisson-fourier}
\end{equation}

Applying the envelope to every electron, $\Gamma=\exp[-\frac12\sum_iw(\bm r_i)]$ maps many-electron states of the lowest Landau level into the AC band.
The AC-band quasihole states are $\Gamma\psi^\sigma_\eta$~\cite{Yan2026}.
Since $\Gamma$ is smooth, positive, and periodic, it preserves the magnetic boundary conditions [Eq.~\eqref{eq:torus-electron-bc}] and the cubic zeros of the Laughlin wave function at coinciding electrons.
For any modulation, $\Gamma\psi^\sigma_\eta$ therefore remains an exact zero mode of the TK interaction, and these states span the $N_\phi$-dimensional space $\Gamma\mathcal H_{\rm qh}$.

\section{Many-body Berry curvature} \label{sec:berry} \label{sec:many-body-berry}

For a family of quasihole states $\ket{\Psi_{\bm k}}$ with one state per total momentum $\bm k$, we use the many-body Berry phase defined by the density form factors~\cite{Geraedts2018},
\begin{equation}
 \Lambda_{\bm q}(\bm k)=\bra{\Psi_{\bm k+\bm q}}\rho_{\bm q}\ket{\Psi_{\bm k}},
 \qquad \rho_{\bm q}=\sum_ie^{\ii\bm q\cdot\bm r_i}.
 \label{eq:density-link}
\end{equation}
With the smallest transfers on the torus, $\bm q_x=(2\pi/L_x)\hat{\bm x}$ and $\bm q_y=(2\pi/L_y)\hat{\bm y}$, the phase around a plaquette defines the Berry curvature and the Chern number,
\begin{equation}
 \Omega(\bm k)=\frac{\gamma_\square(\bm k)}{|\bm q_x||\bm q_y|},\qquad
 \gamma_\square(\bm k)=\operatorname{Arg}\big[
 \Lambda_{\bm q_x}(\bm k)\,\Lambda_{\bm q_y}(\bm k+\bm q_x)\,
 \Lambda_{-\bm q_x}(\bm k+\bm q_x+\bm q_y)\,\Lambda_{-\bm q_y}(\bm k+\bm q_y)\big],
 \qquad
 C=\frac1{2\pi}\sum_{\bm k}\gamma_\square(\bm k).
 \label{eq:curvature-def}
\end{equation}
The sum is an integer whenever all form factors are nonzero~\cite{Fukui2005}.
For several bands, $\ket{\Psi_{\bm k}}$ is replaced by the eigenstates of one band.

The position of the quasihole, the center of its excess charge with the excess electron number $n_{\rm ex}=-1/3$, acts on the amplitudes of a wave packet $\sum_{\bm k}\zeta(\bm k)\ket{\Psi_{\bm k}}$ as (Appendix~B of the main text)
\begin{equation}
 \bm{\mathcal R}=\ii\nabla_{\bm k}+\bm{\mathcal A}(\bm k),\qquad
 [\mathcal R_x,\mathcal R_y]=\ii\Omega(\bm k),
 \label{eq:berry-position}
\end{equation}
where the connection is fixed by the links at small transfer, $\Lambda_{\bm q}(\bm k)=n_{\rm ex}[1+\ii\bm q\cdot\bm{\mathcal A}(\bm k)+O(q^2)]$, and $\Omega=\partial_{k_x}\mathcal A_y-\partial_{k_y}\mathcal A_x$ is the plaquette curvature [Eq.~\eqref{eq:curvature-def}] on a fine mesh.

\subsection{Curvature in the uniform field} \label{sec:qh-berry}

Let $\mathcal P$ project onto the quasihole space without modulation, $\mathcal H_{\rm qh}$, and consider a transfer $\bm q=(2\pi m/L_x,2\pi n/L_y)$ compatible with the torus together with the translation by
\begin{equation}
 \bm d(\bm q)=3\ell_B^2\hat{\bm z}\times\bm q.
 \label{eq:qh-displacement}
\end{equation}
By the translation algebra [Eq.~\eqref{eq:translation-algebra}], $U_{\bm a_j}U_{\bm d(\bm q)}=e^{3\ii N_e\bm q\cdot\bm a_j}U_{\bm d(\bm q)}U_{\bm a_j}=e^{-\ii\bm q\cdot\bm a_j}U_{\bm d(\bm q)}U_{\bm a_j}$, since $3N_e=N_\phi-1$ and $\bm q\cdot\bm a_j\in2\pi\mathbb Z/N_j$.
Thus $U_{\bm d(\bm q)}$ raises the total momentum by $\bm q$, as $\rho_{\bm q}$ does, and since each momentum occurs once,
\begin{equation}
 \mathcal P\rho_{\bm q}\ket{\Psi_{\bm k}}=\Lambda_{\bm q}(\bm k)\ket{\Psi_{\bm k+\bm q}},
 \qquad
 U_{\bm d(\bm q)}\ket{\Psi_{\bm k}}=e^{\ii\alpha_{\bm q}(\bm k)}\ket{\Psi_{\bm k+\bm q}},
 \label{eq:qh-bloch-action}
\end{equation}
with the density link $\Lambda_{\bm q}(\bm k)$ [Eq.~\eqref{eq:density-link}].

We now show that the ratio
$\Lambda_{\bm q}(\bm k)e^{-\ii\alpha_{\bm q}(\bm k)}
=\bra{\Psi_{\bm k+\bm q}}\rho_{\bm q}\ket{\Psi_{\bm k}}/\bra{\Psi_{\bm k+\bm q}}U_{\bm d(\bm q)}\ket{\Psi_{\bm k}}$,
which does not depend on the phases of the Bloch states, is the same at all momenta.
A fine translation $U_{\bm c}$ [Eq.~\eqref{eq:fine-translations}] maps the Bloch states at $\bm k$ and $\bm k+\bm q$ onto those at shifted momenta [Eq.~\eqref{eq:torus-fine-shift}].
Conjugation by it multiplies the numerator and the denominator by the same phase: for the density, by its covariance
\begin{equation}
 U_{\bm c}\rho_{\bm q}U_{\bm c}^{-1}
 =e^{-\ii\bm q\cdot\bm c}\rho_{\bm q},
 \label{eq:rho-covariance}
\end{equation}
and for $U_{\bm d(\bm q)}$, by the translation algebra with $3N_e\bm q\cdot\bm c\equiv-\bm q\cdot\bm c\pmod{2\pi}$.
The ratio therefore takes one value $f(\bm q)$ at all momenta connected by $T_x$ and $T_y$, which are all of them, and the density links of the band without modulation are
\begin{equation}
 \Lambda_{\bm q}(\bm k)=f(\bm q)e^{\ii\alpha_{\bm q}(\bm k)}.
 \label{eq:qh-link}
\end{equation}
Since this holds at every momentum, the projected density is a magnetic translation,
\begin{equation}
 \mathcal P\rho_{\bm q}\mathcal P=f(\bm q)\,U_{\bm d(\bm q)}\mathcal P.
 \label{eq:qh-density}
\end{equation}
Around a plaquette [Eq.~\eqref{eq:curvature-def}] the factors $f$ cancel, and the translation algebra [Eq.~\eqref{eq:translation-algebra}] gives the uniform phase
\begin{equation}
 e^{\ii\gamma_\square}
 =\frac{\bra{\Psi_{\bm k+\bm q_x+\bm q_y}}U_{\bm d(\bm q_y)}U_{\bm d(\bm q_x)}\ket{\Psi_{\bm k}}}
 {\bra{\Psi_{\bm k+\bm q_x+\bm q_y}}U_{\bm d(\bm q_x)}U_{\bm d(\bm q_y)}\ket{\Psi_{\bm k}}}
 =e^{6\pi\ii/N_\phi},
 \label{eq:qh-plaquette}
\end{equation}
which corresponds to the uniform curvature $\Omega=3\ell_B^2$ and $C=3$~\cite{Pu2026}.

\subsection{Curvature in the modulated field} \label{sec:ac-curvature-weak}

The position of the quasihole obeys $[\mathcal R_x,\mathcal R_y]=\ii\Omega(\bm k)$, with $\Omega$ the many-body Berry curvature [Eqs.~\eqref{eq:curvature-def} and \eqref{eq:berry-position}].
Here we compute this position directly for the AC-band quasihole with one flux quantum per cell, and obtain $\Omega$ by comparison.

Consider a single quasihole in the bulk, where we may use the plane and the symmetric gauge.
Its wave function is
\begin{equation}
 \psi_\eta=e^{-|\eta|^2/(4\ell_h^2)}\,
 \prod_j(z_j-\eta)\prod_{i<j}(z_i-z_j)^3\,
 e^{-\sum_j|z_j|^2/(4\ell_B^2)},
 \qquad \ell_h^2=3\ell_B^2,
 \label{eq:ac-planar-qh}
\end{equation}
and with $W=\sum_iw(\bm r_i)$ the weight of the AC-band state is that of a plasma~\cite{Laughlin1983},
\begin{equation}
 \begin{gathered}
 |\psi_\eta|^2e^{-W}=e^{-\frac23U},
 \qquad
 \phi_w(\bm r)=\frac{|\bm r|^2}{4\ell_B^2}+\frac12w(\bm r),\\
 U=-9\sum_{i<j}\ln|z_i-z_j|-3\sum_i\ln|z_i-\eta|
 +3\sum_i\phi_w(\bm r_i)+\frac{|\eta|^2}{4\ell_B^2}.
 \end{gathered}
 \label{eq:ac-plasma}
\end{equation}
The electrons carry the plasma charge $3$, and the quasihole is a fixed impurity of charge $1$.
The potential $\phi_w$ is that of the modulated neutralizing background, the same with and without the quasihole, and the last term of $U$ does not affect the electron density.

Let $n_\eta^w(\bm r)$ be the electron density in the normalized state $\Gamma\ket{\psi_\eta}/\sqrt{\bra{\psi_\eta}\Gamma^2\ket{\psi_\eta}}$, and $n_{\rm bg}^w(\bm r)$ that without a quasihole in the same modulation.
The quasihole changes the electron density by
\begin{equation}
 \delta n_\eta(\bm r)=n_\eta^w(\bm r)-n_{\rm bg}^w(\bm r),
 \label{eq:ac-density-change}
\end{equation}
and the plasma charge density by
\begin{equation}
 \sigma_\eta(\bm r)=\delta^{(2)}(\bm r-\bm\eta)+3\,\delta n_\eta(\bm r),
 \label{eq:ac-screening-charge}
\end{equation}
the impurity together with the change of the electron charge.
Its potential is
\begin{equation}
 u_\eta(\bm r)=-\int\dd^2r'\,\sigma_\eta(\bm r')
 \ln\frac{|\bm r-\bm r'|}{\ell_B},
 \qquad
 \nabla^2u_\eta=-2\pi\sigma_\eta.
 \label{eq:ac-screening-potential}
\end{equation}
We assume that the plasma screens the impurity completely through dipole order: $u_\eta-u_\infty=o(r^{-1})$ and $\nabla u_\eta=o(r^{-2})$ at large $r$, as for exponential screening.
The thermodynamic limit is taken with the quasihole in the bulk before the integrals below, so that the compensating charge at a distant edge is excluded.
For $h=1$, $x$, and $y$, Green's identity on a disk $D_R$ of radius $R$, with the outward normal derivative $\partial_n$, then gives
\begin{equation}
 \int\dd^2r\,h(\bm r)\sigma_\eta(\bm r)
 =-\frac1{2\pi}\lim_{R\to\infty}
 \oint_{\partial D_R}
 \big(h\,\partial_nu_\eta-u_\eta\,\partial_nh\big)\,\dd s=0:
 \label{eq:ac-screening-moments}
\end{equation}
$\sigma_\eta$ has neither a net charge nor a net dipole moment.
With the plasma charge density [Eq.~\eqref{eq:ac-screening-charge}], these are the charge and dipole sum rules~\cite{Gruber1981},
\begin{equation}
 \int\dd^2r\,\delta n_\eta(\bm r)=-\frac13,
 \qquad
 \int\dd^2r\,z\,\delta n_\eta(\bm r)=-\frac\eta3.
 \label{eq:ac-screening-sum-rules}
\end{equation}
The first gives the excess electron number $n_{\rm ex}=-1/3$, and the second places the center of the density deficit at $\eta$, even when the deficit is not symmetric.

We now turn this sum rule into an operator relation for the position of the AC-band quasihole,
\begin{equation}
 \mathcal R_\pm=\mathcal R_x\pm\ii\mathcal R_y
 =\frac1{n_{\rm ex}}P_\Gamma\!\int\dd^2r\,(x\pm\ii y)\,
 \big[\hat n(\bm r)-n_{\rm bg}^w(\bm r)\big]P_\Gamma,
 \label{eq:ac-position-op}
\end{equation}
where $P_\Gamma$ projects onto $\Gamma\mathcal H_{\rm qh}$.
Let $\ket{h_\eta}=e^{|\eta|^2/(4\ell_h^2)}\ket{\psi_\eta}$, whose wave function is holomorphic in $\eta$.
In the state $\Gamma\ket{h_\eta}$, which is proportional to $\Gamma\ket{\psi_\eta}$, the expectation value of $\int\dd^2r\,z\,[\hat n(\bm r)-n_{\rm bg}^w(\bm r)]$ is $\int\dd^2r\,z\,\delta n_\eta=n_{\rm ex}\eta$ by the dipole sum rule [Eq.~\eqref{eq:ac-screening-sum-rules}].
Since $\Gamma\ket{h_\eta}$ lies in $\Gamma\mathcal H_{\rm qh}$, on which $P_\Gamma$ acts as the identity, division by $n_{\rm ex}$ gives
\begin{equation}
 \bra{h_\eta}\Gamma\mathcal R_+\Gamma\ket{h_\eta}
 =\eta\,\bra{h_\eta}\Gamma^2\ket{h_\eta}.
 \label{eq:ac-centroid}
\end{equation}
Neither $\Gamma$ nor $\mathcal R_+$ depends on $\eta$, so that both sides, taken between $\bra{h_{\eta'}}$ and $\ket{h_\eta}$, are antiholomorphic in $\eta'$ and holomorphic in $\eta$.
They agree at $\eta'=\eta$ and hence everywhere,
\begin{equation}
 \bra{h_{\eta'}}\Gamma\mathcal R_+\Gamma\ket{h_\eta}
 =\eta\,\bra{h_{\eta'}}\Gamma^2\ket{h_\eta}.
 \label{eq:ac-position-sum}
\end{equation}
Without modulation, $\Gamma=1$ and $\mathcal R_+$ becomes the position $\mathcal R_+^{(0)}$ without modulation; since the states $\ket{h_\eta}$ span $\mathcal H_{\rm qh}$, the relation then gives
\begin{equation}
 \mathcal R_+^{(0)}\ket{h_\eta}=\eta\ket{h_\eta}.
 \label{eq:ac-coherent}
\end{equation}
Replacing $\eta\ket{h_\eta}$ by $\mathcal R_+^{(0)}\ket{h_\eta}$ on the right-hand side of the relation [Eq.~\eqref{eq:ac-position-sum}] and taking the adjoint, we obtain, between arbitrary states of $\mathcal H_{\rm qh}$,
\begin{equation}
 \mathcal P\Gamma\mathcal R_+\Gamma\mathcal P
 =S\mathcal R_+^{(0)},
 \qquad
 \mathcal P\Gamma\mathcal R_-\Gamma\mathcal P
 =\mathcal R_-^{(0)}S,
 \qquad S=\mathcal P\Gamma^2\mathcal P.
 \label{eq:ac-intertwine}
\end{equation}
Since $\Gamma S^{-1/2}$ maps an orthonormal basis of $\mathcal H_{\rm qh}$ onto an orthonormal basis of $\Gamma\mathcal H_{\rm qh}$, the matrices of $\mathcal R_\pm$ in this basis are
\begin{equation}
 S^{-1/2}\,\mathcal P\Gamma\mathcal R_+\Gamma\mathcal P\,S^{-1/2}
 =S^{1/2}\,\mathcal R_+^{(0)}\,S^{-1/2},
 \qquad
 S^{-1/2}\,\mathcal P\Gamma\mathcal R_-\Gamma\mathcal P\,S^{-1/2}
 =S^{-1/2}\,\mathcal R_-^{(0)}\,S^{1/2}.
 \label{eq:ac-position-general}
\end{equation}
This form also holds with several states per momentum, and we use it for the folded bands in Sec.~\ref{sec:square-magnetic-modulation}.

With one flux quantum per cell, $S$ is diagonal in the Bloch basis, $S\ket{\Psi_{\bm k}}=\mathcal N(\bm k)\ket{\Psi_{\bm k}}$, where
\begin{equation}
 \mathcal N(\bm k)=\langle e^{-W}\rangle_{\bm k},\qquad
 \langle A\rangle_{\bm k}=\bra{\Psi_{\bm k}}A\ket{\Psi_{\bm k}},
 \label{eq:ac-norm}
\end{equation}
is the squared norm of $\Gamma\ket{\Psi_{\bm k}}$, and the orthonormal basis consists of the normalized Bloch states of the AC band.
On their amplitudes, with $\bm{\mathcal R}^{(0)}=\ii\nabla_{\bm k}+\bm{\mathcal A}^{(0)}$ [Eq.~\eqref{eq:berry-position}], the position [Eq.~\eqref{eq:ac-position-general}] becomes
\begin{equation}
 \begin{aligned}
 \mathcal R_+
 &=\mathcal N^{1/2}\mathcal R_+^{(0)}\mathcal N^{-1/2}
 =\mathcal R_+^{(0)}
 -\frac\ii2(\partial_{k_x}+\ii\partial_{k_y})\ln\mathcal N,\\
 \mathcal R_-
 &=\mathcal N^{-1/2}\mathcal R_-^{(0)}\mathcal N^{1/2}
 =\mathcal R_-^{(0)}
 +\frac\ii2(\partial_{k_x}-\ii\partial_{k_y})\ln\mathcal N,
 \end{aligned}
 \label{eq:ac-position-deformed}
\end{equation}
that is, $\bm{\mathcal R}=\ii\nabla_{\bm k}+\bm{\mathcal A}$ with
\begin{equation}
 \mathcal A_x=\mathcal A_x^{(0)}
 +\frac12\partial_{k_y}\ln\mathcal N,
 \qquad
 \mathcal A_y=\mathcal A_y^{(0)}
 -\frac12\partial_{k_x}\ln\mathcal N.
 \label{eq:ac-connection}
\end{equation}
Comparing $[\mathcal R_x,\mathcal R_y]=\ii(\partial_{k_x}\mathcal A_y-\partial_{k_y}\mathcal A_x)$ with $[\mathcal R_x,\mathcal R_y]=\ii\Omega$, and using $\partial_{k_x}\mathcal A^{(0)}_y-\partial_{k_y}\mathcal A^{(0)}_x=3\ell_B^2$ [Eq.~\eqref{eq:qh-plaquette}], we obtain the Berry curvature of the quasihole band,
\begin{equation}
 \Omega(\bm k)=3\ell_B^2
 -\frac12\nabla_{\bm k}^2\ln\mathcal N(\bm k).
 \label{eq:ac-qh-curvature-norm}
\end{equation}
Since $\mathcal N$ is periodic, the correction integrates to zero, and $C=3$.

The same normalization determines the Berry curvature for dragging a pinned quasihole in real space~\cite{Yan2026}, $\mathcal B(\bm\eta)=\frac12\nabla_{\bm\eta}^2\ln\bra{h_\eta}\Gamma^2\ket{h_\eta}$, where $\ket{h_\eta}$ is holomorphic in $\eta$.
Since $S$ is diagonal in the Bloch basis,
\begin{equation}
 \frac{\bra{h_\eta}\Gamma^2\ket{h_\eta}}{\langle h_\eta|h_\eta\rangle}
 =\sum_{\bm k}p_\eta(\bm k)\,\mathcal N(\bm k),
 \qquad
 p_\eta(\bm k)=\frac{|\langle\Psi_{\bm k}|h_\eta\rangle|^2}{\langle h_\eta|h_\eta\rangle},
 \label{eq:ac-real-momentum}
\end{equation}
where $p_\eta$ is the momentum distribution of the localized quasihole, peaked at a momentum set by $\eta$ with a width of order $1/\ell_h$.
The two curvatures thus derive from the same function $\mathcal N$ but probe it differently: $\mathcal B$ describes a quasihole held at a position and averages $\mathcal N$ over momenta, whereas $\Omega$ describes the motion of a quasihole of definite momentum.

For a weak modulation, $\ln\mathcal N=\ln\langle e^{-W}\rangle_{\bm k}$ generates the cumulants of $W$ in the quasihole state without modulation.
By the momentum dependence of periodic operators [Eq.~\eqref{eq:ac-periodic-operator}], the first of them is $-\sum_{\bm G\ne0}w_{\bm G}\langle\rho_{\bm G}\rangle_{\bm 0}\, e^{-\ii\bm k\cdot\bm d(\bm G)}$, and with the harmonics of $w$ [Eq.~\eqref{eq:ac-poisson-fourier}],
\begin{equation}
 \Omega(\bm k)=3\ell_B^2\Big[1+\sum_{\bm G\ne0}\kappa_{\bm G}\,
 \xi(\bm G)\,e^{-\ii\bm k\cdot\bm d(\bm G)}\Big]+O(\kappa^2),
 \qquad
 \xi(\bm G)=3\,\langle\rho_{\bm G}\rangle_{\bm 0}.
 \label{eq:ac-qh-curvature-weak}
\end{equation}
For the first harmonics, $\xi(\pm\bm b_j)=\xi_1$ gives the cosine curvature of the main text.
Unlike the dispersion, its coefficients do not depend on the interaction $V$.
The same expansion gives the first-order correction for the folded bands in Sec.~\ref{sec:square-weak-curvature}, which reduces to this result for one flux quantum per cell, and the Monte Carlo curvature agrees with it without fitted parameters (Fig.~3 of the main text).

\section{Quasihole bands in a periodic modulation} \label{sec:square-magnetic-modulation}

The space $\Gamma\mathcal H_{\rm qh}$ is exactly degenerate.
The interaction $\alpha V_{\rm TK}+V$, with $\alpha\to\infty$ and $V$ weak compared with the cyclotron gap, lifts the degeneracy, and the quasihole states are the eigenstates of $V$ projected onto $\Gamma\mathcal H_{\rm qh}$.
If the cell of the modulation encloses $p$ flux quanta, the $p$ Bloch states $\ket{\Psi_{\bm k,\alpha}}$, $\alpha=1,\ldots,p$, whose momenta differ by reciprocal vectors of the modulation have the same crystal momentum.
The states $\Gamma\ket{\Psi_{\bm k,\alpha}}$ are not orthogonal, and the bands $E_n$ and the eigenstates $\Gamma\sum_\alpha c_{n,\alpha}\ket{\Psi_{\bm k,\alpha}}$ follow from the generalized eigenvalue problem
\begin{equation}
 H_{\alpha\beta}=\bra{\Psi_{\bm k,\alpha}}\Gamma V\Gamma\ket{\Psi_{\bm k,\beta}},
 \quad S_{\alpha\beta}=\bra{\Psi_{\bm k,\alpha}}\Gamma^2\ket{\Psi_{\bm k,\beta}},
 \quad Hc_n=E_nSc_n,\quad c_n^\dagger Sc_n=1.
 \label{eq:ac-projected-problem}
\end{equation}
For a rational flux $p/q$ per cell, the same holds for the magnetic cell of $q$ cells, which encloses $p$ flux quanta (Sec.~\ref{sec:chern-diophantine}).

\subsection{Dispersion of the single band} \label{sec:ac-dispersion}

For $p=1$, $H$ and $S$ are numbers.
With $W=\sum_{\bm G\ne0}w_{\bm G}\rho_{\bm G}$, the energy $E=\langle Ve^{-W}\rangle_{\bm k}/\langle e^{-W}\rangle_{\bm k}$ expands as
\begin{equation}
 E(\bm k)=E_0-\sum_{\bm G\ne0}w_{\bm G}
 \big[\langle V\rho_{\bm G}\rangle_{\bm k}
 -E_0\langle\rho_{\bm G}\rangle_{\bm k}\big]+O(w^2),
 \label{eq:ac-first-order}
\end{equation}
where $\langle V\rangle_{\bm k}=E_0$ because $V$ commutes with the fine translations, which connect all momenta.
The same translations fix the momentum dependence [Eqs.~\eqref{eq:torus-fine-shift} and \eqref{eq:rho-covariance}]: an operator $O_{\bm G}$ that transforms like $\rho_{\bm G}$, with $\bm G=m_1\bm b_1+m_2\bm b_2$, satisfies
\begin{equation}
 \langle O_{\bm G}\rangle_{\bm k}
 =e^{-\ii\bm k\cdot\bm d(\bm G)}\langle O_{\bm G}\rangle_{\bm0},
 \qquad \bm d(\bm G)=3(m_1\bm a_2-m_2\bm a_1).
 \label{eq:ac-periodic-operator}
\end{equation}
Hence
\begin{equation}
 \chi(\bm G)=\langle(V-E_0)\rho_{\bm G}\rangle_{\bm0},\qquad
 E(\bm k)=E_0-\sum_{\bm G\ne0}w_{\bm G}\chi(\bm G)
 e^{-\ii\bm k\cdot\bm d(\bm G)}+O(w^2),
 \label{eq:ac-dispersion-general}
\end{equation}
with $\chi$ real and even by Hermiticity and inversion.
Since $V$ commutes with the magnetic translations, the argument of Sec.~\ref{sec:qh-berry} applies to $(V-E_0)\rho_{\bm G}$ and gives the operator form
\begin{equation}
 \mathcal P(V-E_0)\rho_{\bm G}\mathcal P
 =\chi(\bm G)U_{\bm a_2}^{3m_1}U_{\bm a_1}^{-3m_2}\mathcal P.
 \label{eq:ac-chi-operator}
\end{equation}
For the first harmonics of the main text, $w_{\pm\bm b_1}=-\kappa_x/(2\pi)$, $w_{\pm\bm b_2}=-\kappa_y/(2\pi)$, and $\chi(\pm\bm b_j)=\chi_1$ by square symmetry, which gives the cosine dispersion of the main text.
Applied to products of harmonics, the same momentum dependence [Eq.~\eqref{eq:ac-periodic-operator}] gives the periods $\bm b_1/3$ and $\bm b_2/3$ to all orders in $w$~\cite{Yan2026,Iyer2026}.

\subsection{Band folding at two flux quanta per cell} \label{sec:square-folding} \label{sec:square-weak}

For the square modulation of the main text, the cell is spanned by $\tilde{\bm a}_1=(a,a)$ and $\tilde{\bm a}_2=(-a,a)$, with the reciprocal vectors $\tilde{\bm b}_1=(\pi/a)(1,1)$ and $\tilde{\bm b}_2=(\pi/a)(-1,1)$, and the reduced zone is $|k_x\pm k_y|\leq\pi/a$.
We fix the relative phase of the two states at each momentum by
\begin{equation}
 \ket{\Psi_{\bm k,1}}=\ket{\Psi_{\bm k}},\qquad
 \ket{\Psi_{\bm k,2}}=U_{\bm d(\tilde{\bm b}_1)}\ket{\Psi_{\bm k}}
 \propto\ket{\Psi_{\bm k+\tilde{\bm b}_1}},
 \label{eq:square-reference-pair}
\end{equation}
and work in the orthonormal basis,
\begin{equation}
 H_{\rm eff}=S^{-1/2}HS^{-1/2},\qquad
 H_{\rm eff}v_n=E_nv_n,\qquad c_n=S^{-1/2}v_n,
 \quad v_m^\dagger v_n=\delta_{mn}.
 \label{eq:square-horth}
\end{equation}
The modulation has the harmonics
\begin{equation}
 w_{\bm G}=-\frac\kappa{2\pi}\quad\text{for }\bm G=\pm\bm b_1,\pm\bm b_2,
 \qquad
 w_{\bm G}=-\frac{\tilde\kappa}\pi\quad\text{for }\bm G=\pm\tilde{\bm b}_1,\pm\tilde{\bm b}_2,
 \label{eq:square-w}
\end{equation}
and with $\Gamma V\Gamma=Ve^{-W}$ and $\Gamma^2=1-W+O(w^2)$,
\begin{equation}
 \begin{aligned}
 S_{\alpha\beta}(\bm k)&=\delta_{\alpha\beta}
 -\mathcal W_{\alpha\beta}(\bm k)+O(w^2),
 \qquad
 \mathcal W_{\alpha\beta}(\bm k)=\sum_{\bm G}w_{\bm G}\,
 \bra{\Psi_{\bm k,\alpha}}\rho_{\bm G}\ket{\Psi_{\bm k,\beta}},\\
 H_{\alpha\beta}(\bm k)&=E_0\,\delta_{\alpha\beta}
 -\sum_{\bm G}w_{\bm G}\,
 \bra{\Psi_{\bm k,\alpha}}V\rho_{\bm G}\ket{\Psi_{\bm k,\beta}}+O(w^2).
 \end{aligned}
 \label{eq:square-first-order-matrices}
\end{equation}
Since $S^{-1/2}=1+\mathcal W/2+O(w^2)$, the overlap replaces $V$ by $V-E_0$,
\begin{equation}
 \big[H_{\rm eff}(\bm k)\big]_{\alpha\beta}
 =E_0\,\delta_{\alpha\beta}-\sum_{\bm G}w_{\bm G}\,
 \bra{\Psi_{\bm k,\alpha}}(V-E_0)\rho_{\bm G}\ket{\Psi_{\bm k,\beta}}
 +O(w^2).
 \label{eq:square-horth-first-order}
\end{equation}
The two states are degenerate without modulation, so the splitting is of first order in $w$, while the eigenvectors, and with them the curvature, are of order one.

Since $V$ commutes with the magnetic translations, the argument of Sec.~\ref{sec:qh-berry} shows that $(V-E_0)\rho_{\bm G}$ acts between states without modulation as a magnetic translation for every $\bm G$ compatible with the torus,
\begin{equation}
 \bra{\Psi_{\bm k'}}(V-E_0)\rho_{\bm G}\ket{\Psi_{\bm k}}
 =\chi(\bm G)\,\bra{\Psi_{\bm k'}}U_{\bm d(\bm G)}\ket{\Psi_{\bm k}},
 \label{eq:square-chi-operator}
\end{equation}
with the real and even coefficient
\begin{equation}
 \chi(\bm G)=\bra{\Psi_{\bm G}}V\rho_{\bm G}\ket{\Psi_{\bm 0}}
 -\langle V\rangle_{\bm 0}\,\bra{\Psi_{\bm G}}\rho_{\bm G}\ket{\Psi_{\bm 0}},
 \qquad \ket{\Psi_{\bm G}}=U_{\bm d(\bm G)}\ket{\Psi_{\bm0}}.
 \label{eq:square-chi-covariance}
\end{equation}
Fourfold symmetry leaves $\chi_1=\chi(\pm\bm b_j)$ and $\tilde\chi=\chi(\pm\tilde{\bm b}_j)$, and with the translation matrices
\begin{equation}
 \big[\mathcal U_{\bm G}(\bm k)\big]_{\alpha\beta}
 =\bra{\Psi_{\bm k,\alpha}}U_{\bm d(\bm G)}\ket{\Psi_{\bm k,\beta}},
 \qquad
 \mathcal U_{-\bm G}=\mathcal U_{\bm G}^\dagger,
 \label{eq:square-translation-block}
\end{equation}
we obtain $H_{\rm eff}=E_0+H^{(1)}+O(w^2)$ with
\begin{equation}
 H^{(1)}(\bm k)=
 \frac{\kappa\chi_1}{2\pi}\sum_{\bm G=\pm\bm b_1,\pm\bm b_2}\mathcal U_{\bm G}(\bm k)
 +\frac{\tilde\kappa\tilde\chi}{\pi}
 \sum_{\bm G=\pm\tilde{\bm b}_1,\pm\tilde{\bm b}_2}\mathcal U_{\bm G}(\bm k).
 \label{eq:square-heff}
\end{equation}

The translations at $\bm b_j$ are diagonal in the folded basis, and those at $\tilde{\bm b}_j$ exchange its two states; since $3\ell_B^2(\tilde{\bm b}_1\times\tilde{\bm b}_2)_z=3\pi$, the latter anticommute.
With $\theta_\pm=(3a/2)(k_x\pm k_y)$, the translation algebra [Eq.~\eqref{eq:translation-algebra}] gives
\begin{equation}
 \begin{aligned}
 \mathcal U_{\bm b_1}(\bm k)&=e^{-\ii(\theta_+-\theta_-)}\,\sigma_z,
 &\qquad
 \mathcal U_{\bm b_2}(\bm k)&=e^{\ii(\theta_++\theta_-)}\,\sigma_z,\\
 \mathcal U_{\tilde{\bm b}_1}(\bm k)&=
 \begin{pmatrix}0&-e^{2\ii\theta_-}\\1&0\end{pmatrix},
 &\qquad
 \mathcal U_{\tilde{\bm b}_2}(\bm k)&=-\ii\,e^{\ii\theta_+}
 \begin{pmatrix}0&e^{\ii\theta_-}\\e^{-\ii\theta_-}&0\end{pmatrix},
 \end{aligned}
 \label{eq:square-translation-matrices}
\end{equation}
where the first column of $\mathcal U_{\tilde{\bm b}_1}$ is the definition of the folded basis and its square is the lattice translation $U_{2\bm d(\tilde{\bm b}_1)}$.
Hence
\begin{equation}
 H^{(1)}(\bm k)=
 \begin{pmatrix}\varepsilon(\bm k)&\Delta(\bm k)^*\\
 \Delta(\bm k)&-\varepsilon(\bm k)\end{pmatrix},
 \qquad
 \begin{aligned}
 \varepsilon(\bm k)&=\frac{2\kappa\chi_1}{\pi}\cos\theta_+\cos\theta_-,\\
 \Delta(\bm k)&=\frac{2\tilde\kappa\tilde\chi}{\pi}\,e^{-\ii\theta_-}
 \big(\sin\theta_++\ii\sin\theta_-\big),
 \end{aligned}
 \label{eq:square-block}
\end{equation}
and the two bands
\begin{equation}
 E_\pm=E_0\pm\sqrt{\varepsilon^2+|\Delta|^2}
 \label{eq:square-bands}
\end{equation}
are separated by the direct gap $(4/\pi)\min(|\kappa\chi_1|,|\tilde\kappa\tilde\chi|)$.
At second order, the coefficients $\kappa\chi_1$ and $\tilde\kappa\tilde\chi$ are renormalized, both bands are shifted by a common function of $\bm k$, and new harmonics enter $H^{(1)}$.

\subsection{Curvature and Chern numbers of the folded bands} \label{sec:square-weak-curvature}

The density link of band $n$ between $\bm k$ and $\bm k+\bm q_\mu$, $\mu=x,y$, is
\begin{equation}
 \Lambda_{n,\bm q_\mu}(\bm k)
 =c_n^\dagger(\bm k+\bm q_\mu)\,D_{\bm q_\mu}(\bm k)\,c_n(\bm k),
 \qquad
 \big[D_{\bm q}(\bm k)\big]_{\alpha\beta}
 =\bra{\Psi_{\bm k+\bm q,\alpha}}\Gamma^2\rho_{\bm q}\ket{\Psi_{\bm k,\beta}}.
 \label{eq:square-link-matrix}
\end{equation}
At leading order $\Gamma^2\to1$ and $S\to1$, so that $c_n$ is the eigenvector of $H^{(1)}$ [Eq.~\eqref{eq:square-block}], and by the projected density [Eq.~\eqref{eq:qh-density}] $D_{\bm q}$ is $f(\bm q)$ times the matrix of $U_{\bm d(\bm q)}$.
This translation maps $\ket{\Psi_{\bm k,1}}$ to $e^{\ii\alpha_{\bm q}(\bm k)}\ket{\Psi_{\bm k+\bm q,1}}$ [Eq.~\eqref{eq:qh-link}], and $\ket{\Psi_{\bm k,2}}=U_{\bm d(\tilde{\bm b}_1)}\ket{\Psi_{\bm k,1}}$ to the same state times $e^{-\ii\varphi_\mu}$, the phase from commuting it through $U_{\bm d(\tilde{\bm b}_1)}$, with $\varphi_\mu=3\ell_B^2(\bm q_\mu\times\tilde{\bm b}_1)_z=\theta_-(\bm k+\bm q_\mu)-\theta_-(\bm k)$.
The rotation $\tilde c_n=e^{-\ii\theta_-\sigma_z/2}c_n$ removes this relative phase,
\begin{equation}
 \Lambda_{n,\bm q_\mu}(\bm k)=f(\bm q_\mu)\,
 e^{\ii\alpha_{\bm q_\mu}(\bm k)}\,e^{-\ii\varphi_\mu/2}\;
 \tilde c_n^\dagger(\bm k+\bm q_\mu)\,\tilde c_n(\bm k),
 \label{eq:square-rotation}
\end{equation}
and around a plaquette the scalar factors leave the uniform phase [Eq.~\eqref{eq:qh-plaquette}],
\begin{equation}
 \gamma_{\square,n}=\frac{6\pi}{N_\phi}
 +\operatorname{Arg}\prod_{j=0}^{3}
 \tilde c_n^\dagger(\bm k_{j+1})\tilde c_n(\bm k_j)
 \pmod{2\pi},
 \label{eq:square-plaquette-split}
\end{equation}
where $\bm k_j$ run counterclockwise around the plaquette and $\bm k_4=\bm k_0$.
In the rotated frame, $\tilde c_n$ is an eigenvector of $\bm h\cdot\bm\sigma$ with $h_x+\ii h_y=e^{\ii\theta_-}\Delta$ and $h_z=\varepsilon$, the vector $\bm h$ of the main text, so that
\begin{equation}
 \Omega^{(0)}_\mp=3\ell_B^2\pm\tfrac12\hat{\bm h}\cdot
 (\partial_{k_x}\hat{\bm h}\times\partial_{k_y}\hat{\bm h}),
 \label{eq:square-curvature-general}
\end{equation}
with the upper sign for the lower band.
With $\partial(\theta_+,\theta_-)/\partial(k_x,k_y)=-9\pi\ell_B^2$,
\begin{equation}
 \Omega^{(0)}_\mp(\bm k)=3\ell_B^2\left[1\mp\frac{3\pi}{2}\,
 \frac{\tilde\kappa^2\tilde\chi^2\;\kappa\chi_1\,\big(1-\sin^2\theta_+\sin^2\theta_-\big)}
 {\big[\kappa^2\chi_1^2\cos^2\theta_+\cos^2\theta_-
 +\tilde\kappa^2\tilde\chi^2(\sin^2\theta_++\sin^2\theta_-)\big]^{3/2}}\right].
 \label{eq:square-curvature}
\end{equation}

Since $\bm h(\bm k+\tilde{\bm b}_1)=(-h_x,h_y,-h_z)$, $\bm h$ is periodic only on the zone without folding, which covers the reduced zone twice.
On that zone the uniform term integrates to $3$, and $\hat{\bm h}$ covers the sphere once in each of the $3\times3$ periods of the bands, with the orientation $-\operatorname{sgn}(\kappa\chi_1)$, which gives $-9\operatorname{sgn}(\kappa\chi_1)$.
The curvature of each band is invariant under $\bm k\to\bm k+\tilde{\bm b}_1$, so the reduced zone carries half of each integral,
\begin{equation}
 \begin{aligned}
 C_\mp&=\frac32\mp\frac92\,\operatorname{sgn}(\kappa\chi_1),\\[4pt]
 (C_-,C_+)&=\begin{cases}
 (+6,-3),&\kappa\chi_1<0,\\
 (-3,+6),&\kappa\chi_1>0.
 \end{cases}
 \end{aligned}
 \label{eq:square-cherns-weak}
\end{equation}
Neither term is an integer by itself, but their sum is, and $C_-+C_+=3$.

At first order in the modulation the curvature changes in two ways.
The deformation of the links by $\Gamma$ and $S$ follows from the expansion of the position operator [Eq.~\eqref{eq:ac-position-general}]: the link of band $n$ is multiplied by $1-\sum_{\bm G}w_{\bm G}\,\tfrac12\ell_B^2\xi(\bm G)\,(\bm q\cdot\bm G)\,\langle U_{\bm d(\bm G)}\rangle_{n,\bm k}$, with $\xi(\bm G)=3f(\bm G)$ and $\langle U_{\bm d(\bm G)}\rangle_{n,\bm k}=v_n^\dagger\,\mathcal U_{\bm G}\,v_n$.
With $\xi_1=\xi(\bm b_j)$ and $\tilde\xi=\xi(\tilde{\bm b}_j)$, the translation matrices [Eq.~\eqref{eq:square-translation-matrices}] give
\begin{equation}
 \begin{aligned}
 \delta\Omega^{\rm link}_\mp(\bm k)=\pm3\ell_B^2\,\Big\{
 &\tilde\kappa\,\tilde\xi\big[\partial_{\theta_+}\big(\cos\theta_+\,\hat h_x\big)
 +\partial_{\theta_-}\big(\cos\theta_-\,\hat h_y\big)\big]\\
 &-\kappa\,\xi_1\big[\partial_{\theta_+}\big(\sin\theta_+\cos\theta_-\,\hat h_z\big)
 +\partial_{\theta_-}\big(\cos\theta_+\sin\theta_-\,\hat h_z\big)\big]\Big\}.
 \end{aligned}
 \label{eq:square-link-correction}
\end{equation}
This is a total derivative of a periodic function and does not change the Chern numbers; for $\tilde\kappa\to0$ it reduces to the single-band curvature [Eq.~\eqref{eq:ac-qh-curvature-weak}].
The second-order part of $H_{\rm eff}$ also changes the eigenvectors at relative order $w$, which contributes to the curvature at the same order and is not fixed by $\chi$ and $\xi$.

The leading-order continuous curvature [Eq.~\eqref{eq:square-curvature}] captures the momentum dependence of the Monte Carlo curvature [Fig.~1 of the main text], and the Monte Carlo bands have $(C_-,C_+)=(6,-3)$ in every bootstrap resample, as do those at $N_\phi=400$ and at $d=5\ell_B$.

\subsection{Triangular modulation} \label{sec:triangular-c6}

The splitting into Chern bands of opposite sign does not rely on the square lattice.
Consider a triangular cell spanned by $\bm A_1$ and $\bm A_2$, of equal length and at $60^\circ$, with $(\bm A_1\times\bm A_2)_z=4\pi\ell_B^2$, i.e., two flux quanta, and the field
\begin{equation}
 B_z(\bm r)=-B_0\Big[1+\tilde\kappa\sum_{j=1}^3\cos(\bm G_j\cdot\bm r)
 +\kappa\sum_{j=1}^3\cos(\bm K_j\cdot\bm r)\Big],
 \label{eq:tri-field}
\end{equation}
where $\bm A_i\cdot\bm G_j=2\pi\delta_{ij}$, $\bm G_3=\bm G_1+\bm G_2$, and $\bm K_1=\bm G_1-\bm G_2$, $\bm K_2=\bm G_1+2\bm G_2$, $\bm K_3=2\bm G_1+\bm G_2$ are the two shortest stars of reciprocal vectors.
The harmonics of $w$ [Eq.~\eqref{eq:ac-poisson-fourier}] are $w_{\pm\bm G_j}=-\sqrt3\tilde\kappa/2\pi$ and $w_{\pm\bm K_j}=-\kappa/2\sqrt3\pi$, and the coefficients $\chi$ [Eq.~\eqref{eq:square-chi-covariance}] are equal within each star, so that, as for the square modulation,
\begin{equation}
 H^{(1)}=\sum_{j=1}^3
 \Big[\frac{\sqrt3\,\tilde\kappa\chi(\bm G_1)}{2\pi}\,\mathcal U_{\bm G_j}+\frac{\kappa\chi(\bm K_1)}{2\sqrt3\,\pi}\,\mathcal U_{\bm K_j}+\mathrm{H.c.}\Big],
 \label{eq:tri-response}
\end{equation}
Since $3\ell_B^2(\bm G_1\times\bm G_2)_z=3\pi$, the translations $\mathcal U_{\bm G_1}$ and $\mathcal U_{\bm G_2}$ anticommute, and with $\theta_1=-\tfrac32\bm k\cdot\bm A_2$ and $\theta_2=\tfrac32\bm k\cdot\bm A_1$ we can take $\mathcal U_{\bm G_1}=e^{\ii\theta_1}\sigma_x$ and $\mathcal U_{\bm G_2}=e^{\ii\theta_2}\sigma_z$, which fixes the other harmonics.
Then $H^{(1)}=\bm h\cdot\bm\sigma$ with
\begin{equation}
 \begin{aligned}
 h_x&=\frac{1}{\sqrt3\,\pi}\big[3\tilde\kappa\chi(\bm G_1)\cos\theta_1-\kappa\chi(\bm K_1)\cos(\theta_1+2\theta_2)\big],\\
 h_y&=\frac{1}{\sqrt3\,\pi}\big[-3\tilde\kappa\chi(\bm G_1)\cos(\theta_1+\theta_2)+\kappa\chi(\bm K_1)\cos(\theta_1-\theta_2)\big],\\
 h_z&=\frac{1}{\sqrt3\,\pi}\big[3\tilde\kappa\chi(\bm G_1)\cos\theta_2-\kappa\chi(\bm K_1)\cos(2\theta_1+\theta_2)\big].
 \end{aligned}
 \label{eq:tri-dvector}
\end{equation}
The curvature is given by the formula for the square modulation [Eq.~\eqref{eq:square-curvature-general}] with this $\bm h$.
Over the reduced zone the uniform term contributes $3/2$ to each band and the texture $\mp9/2$, so that
\begin{equation}
 (C_-,C_+)=(-3,6)\quad\text{for }\tilde\kappa\chi(\bm G_1)>0,
 \label{eq:tri-chern}
\end{equation}
and the two bands are exchanged for $\tilde\kappa\chi(\bm G_1)<0$.
The gap stays open, and these Chern numbers persist, as long as $|\kappa\chi(\bm K_1)|<\tfrac32|\tilde\kappa\chi(\bm G_1)|$.

\begin{figure}[!htbp]
 \centering
 \includegraphics[width=\textwidth]{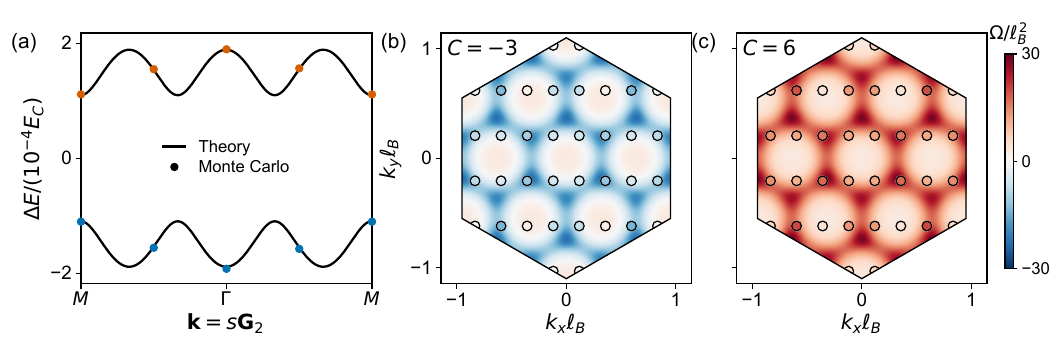}
 \caption{Triangular modulation at $(\tilde\kappa,\kappa)=(0.05,0.05)$, $(N_\phi,N_e)=(64,21)$, and $d=\ell_B$.
(a) Energies $\Delta E=E-\overline E$ along $\bm k=s\bm G_2$, with $\overline E$ the mean over both bands and momenta; black curves show first-order theory and colored points show Monte Carlo data, with one bootstrap standard deviation as error bars.
(b),(c) Lower- and upper-band density-link curvature; continuous first-order theory is shown as the background and finite-plaquette Monte Carlo estimates as circles at the plaquette centers, on the same color scale.
Theory uses the uniform-state covariance coefficients measured in the same sample.
The hexagon is the reduced Brillouin zone; periodic boundary points and the two equivalent $M$ endpoints are repeated.}
 \label{fig:tri-c6-numerics}
\end{figure}
\FloatBarrier

The microscopic calculation at $(\tilde\kappa,\kappa)=(0.05,0.05)$, $(N_\phi,N_e)=(64,21)$, and $d=\ell_B$ agrees with the first-order dispersion, whose coefficients are measured in the state without modulation, and gives $(C_-,C_+)=(-3,6)$ in every bootstrap resample [Fig.~\ref{fig:tri-c6-numerics}].

\subsection{Periodic scalar potential} \label{sec:periodic-electric-potential}

\begin{figure}[!htbp]
 \centering
 \includegraphics[width=\textwidth]{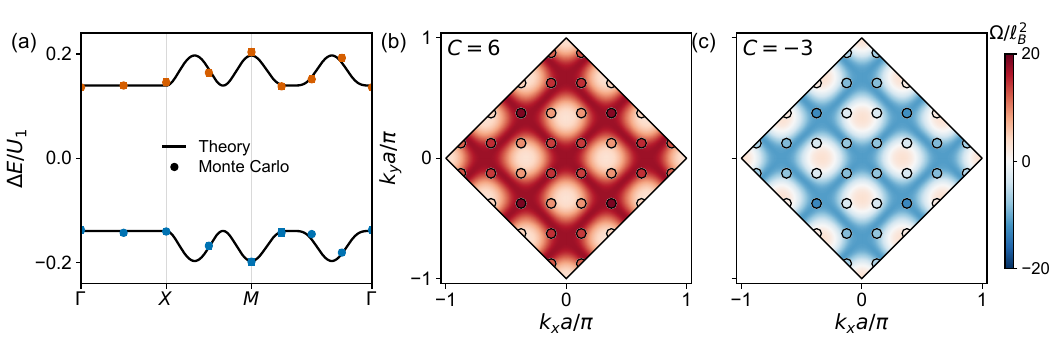}
 \caption{Quasihole bands induced by the periodic electric potential [Eq.~\eqref{eq:electric-periodic-potential}], with $U_1=1$ as the energy unit, $U_2=0.323$, and Monte Carlo data at $(N_\phi,N_e)=(64,21)$.
(a) Energies $\Delta E=E-\overline E$, with $\overline E$ the mean over both bands and momenta; curves show Eq.~\eqref{eq:electric-two-band}.
(b),(c) Lower- and upper-band density-link curvature; backgrounds show Eq.~\eqref{eq:electric-curvature}.
Points are Monte Carlo data projected onto $\mathcal H_{\rm qh}$; energy error bars denote one bootstrap standard deviation.
Theory uses independently measured uniform-state form factors at $N_\phi=100$.
The curvature points lie at plaquette centers, with periodic boundary points repeated.}
 \label{fig:electric-periodic-bands}
\end{figure}

A weak periodic electric potential can also generate quasihole dispersion in the uniform field $B_z=-B_0$.
On the square lattice of Sec.~\ref{sec:square-folding}, we take the potential energy
\begin{equation}
 \hat V_{\rm el}=\sum_i v_{\rm el}(\bm r_i),\qquad
 v_{\rm el}(\bm r)=U_1\sum_{j=1}^2\cos(\bm b_j\cdot\bm r)
 +U_2\sum_{j=1}^2\cos(\tilde{\bm b}_j\cdot\bm r),
 \label{eq:electric-periodic-potential}
\end{equation}
whose cell encloses two flux quanta.
By the projected density [Eq.~\eqref{eq:qh-density}], its first-order Hamiltonian in the folded basis is
\begin{equation}
 H_{\rm el}^{(1)}(\bm k)
 =\frac{U_1f_1}{2}\sum_{\bm G=\pm\bm b_1,\pm\bm b_2}\mathcal U_{\bm G}(\bm k)
 +\frac{U_2f_2}{2}\sum_{\bm G=\pm\tilde{\bm b}_1,\pm\tilde{\bm b}_2}\mathcal U_{\bm G}(\bm k),
 \label{eq:electric-projection}
\end{equation}
with the density form factors $f_1=f(\bm b_j)$ and $f_2=f(\tilde{\bm b}_j)$ of the state without modulation.
In the rotated frame [Eq.~\eqref{eq:square-rotation}],
\begin{equation}
 H_{\rm el}^{(1)}=\bm h_{\rm el}\cdot\bm\sigma,\qquad
 \bm h_{\rm el}=
 \bigl(J_2\sin\theta_+,\ J_2\sin\theta_-,\ J_1\cos\theta_+\cos\theta_-\bigr),\qquad
 J_1=2U_1f_1,\quad J_2=U_2f_2,
 \label{eq:electric-two-band}
\end{equation}
so that the gap is $2\min(|J_1|,|J_2|)$ and both harmonics are needed to open it.
As for the magnetic modulation [Eq.~\eqref{eq:square-curvature-general}],
\begin{equation}
 \Omega_\mp^{\rm el}(\bm k)=3\ell_B^2
 \mp\frac{9\pi\ell_B^2}{2}
 \frac{J_1J_2^2\bigl(1-\sin^2\theta_+\sin^2\theta_-\bigr)}{|\bm h_{\rm el}|^3},
 \qquad
 C_\mp=\frac32\mp\frac92\operatorname{sgn}J_1,
 \label{eq:electric-curvature}
\end{equation}
so that $(C_-,C_+)=(6,-3)$ for $J_1<0$, reversed for $J_1>0$, and independent of the sign of $J_2$ [Fig.~\ref{fig:electric-periodic-bands}].
As for the magnetic modulation, these results hold at first order and neglect Landau-level mixing.

\FloatBarrier

For Fig.~\ref{fig:electric-periodic-bands}, we take $U_1=1$ as the energy unit and $U_2=0.323$.
At fixed $U_2/U_1$, changing $U_1$ only rescales the projected energies.
The Monte Carlo calculation at $(N_\phi,N_e)=(64,21)$ agrees with the first-order dispersion [Eq.~\eqref{eq:electric-two-band}] and gives $(C_-,C_+)=(6,-3)$ in all bootstrap resamples.

\section{Chern numbers from the magnetic translation algebra} \label{sec:chern-diophantine}

With two flux quanta per cell, the square and the triangular magnetic modulations and the periodic potential all give $(C_-,C_+)=(6,-3)$ or $(-3,6)$.
We show that at weak modulation the magnetic translation algebra restricts the Chern number of every isolated band, for any modulation with a rational flux per cell, and that these are the smallest values allowed at two flux quanta per cell.

Let $\bm A_1,\bm A_2$ span the cell of the modulation, with $(\bm A_1\times\bm A_2)_z=2\pi(p/q)\ell_B^2$ for coprime $p$ and $q$, and let $\bm G_i\cdot\bm A_j=2\pi\delta_{ij}$.
The translations $U_{\bm A_1}$ and $U_{\bm A_2}$ commute with $\Gamma$ and $V$, but $U_{\bm A_1}U_{\bm A_2}=e^{2\pi\ii N_ep/q}\,U_{\bm A_2}U_{\bm A_1}$ [Eq.~\eqref{eq:translation-algebra}], so for $q>1$ they do not commute with each other in general.
The translations $U_{q\bm A_1}$ and $U_{\bm A_2}$ do commute, and we label the states by the momentum $\bm k$, with eigenvalues $e^{-\ii\bm k\cdot q\bm A_1}$ and $e^{-\ii\bm k\cdot\bm A_2}$.
Their cell, the magnetic cell, consists of $q$ cells of the modulation and encloses $p$ flux quanta.
The torus contains an integer number of magnetic cells, so $p$ divides $N_\phi=3N_e+1$ and is coprime to $3$.
Each momentum $\bm k$ then holds $p$ states $\ket{\Psi_{\bm k,\alpha}}$, and the generalized eigenvalue problem [Eq.~\eqref{eq:ac-projected-problem}] gives $p$ bands.
For $w(\bm r)=\sum_{\bm G}w_{\bm G}e^{\ii\bm G\cdot\bm r}$ with $\bm G=n_1\bm G_1+n_2\bm G_2$, the first-order Hamiltonian follows as for the square modulation [Eqs.~\eqref{eq:square-horth-first-order} and \eqref{eq:square-chi-operator}],
\begin{equation}
 H^{(1)}_{\alpha\beta}(\bm k)=-\sum_{\bm G}w_{\bm G}\,\chi(\bm G)\,
 \bra{\Psi_{\bm k,\alpha}}U_{\bm d(\bm G)}\ket{\Psi_{\bm k,\beta}},
 \qquad \bm d(\bm G)=3\ell_B^2\hat{\bm z}\times\bm G.
 \label{eq:chern-first-order}
\end{equation}
All translations in this sum are products of $X=U_{\bm d(\bm G_1)}=U_{(3q/p)\bm A_2}$ and $Y=U_{\bm d(\bm G_2)}=U_{-(3q/p)\bm A_1}$, $U_{\bm d(\bm G)}=\eta_{\bm G}X^{n_1}Y^{n_2}$ with a momentum-independent phase $\eta_{\bm G}$.
With $3N_e=N_\phi-1$, the translation algebra gives
\begin{equation}
 XY=\omega\,YX,\qquad \omega=e^{-6\pi\ii q/p},\qquad
 X^p=e^{-3\ii q\bm k\cdot\bm A_2},\qquad Y^p=e^{3\ii q\bm k\cdot\bm A_1}.
 \label{eq:chern-algebra}
\end{equation}
Since $p$ is coprime to $3q$, $\omega$ is a primitive $p$th root of unity, and the $p$ states at $\bm k$ have a basis $\ket m$, $m=0,\ldots,p-1$, with
\begin{equation}
 X\ket m=e^{-(3\ii q/p)\bm k\cdot\bm A_2}\,\omega^m\ket m,\qquad
 Y\ket m=e^{(3\ii q/p)\bm k\cdot\bm A_1}\ket{m+1},\qquad \ket p\equiv\ket0.
 \label{eq:chern-matrices}
\end{equation}
In this basis,
\begin{equation}
 H^{(1)}(\bm k)=\sum_{n_1,n_2}h_{n_1n_2}\,
 e^{(3\ii q/p)\bm k\cdot(n_2\bm A_1-n_1\bm A_2)}\,Z^{n_1}P^{n_2},
 \qquad h_{n_1n_2}=-w_{\bm G}\chi(\bm G)\eta_{\bm G},
 \label{eq:chern-fourier}
\end{equation}
with the clock and shift matrices $Z=\operatorname{diag}(1,\omega,\ldots,\omega^{p-1})$ and $P\ket m=\ket{m+1}$, which obey $ZP=\omega PZ$.
This is the Hamiltonian of a particle in a Landau level with a weak periodic potential and $p/(3q)$ flux quanta per cell~\cite{Thouless1982}, since the charge $e/3$ sees one third of the flux.

The Chern number follows from the density links, as for the square modulation (Sec.~\ref{sec:square-weak-curvature}).
For a small transfer $\bm q$, the translation $U_{\bm d(\bm q)}$ commutes with $X$ and $Y$ up to the phases $e^{-3\ii q\,\bm q\cdot\bm A_2/p}$ and $e^{3\ii q\,\bm q\cdot\bm A_1/p}$, which are the changes of the prefactors of $X\ket m$ and $Y\ket m$ under $\bm k\to\bm k+\bm q$ [Eq.~\eqref{eq:chern-matrices}].
It therefore maps $\ket m$ at $\bm k$ to $\ket m$ at $\bm k+\bm q$, up to a phase $e^{\ii\alpha_{\bm q}(\bm k)}$ common to all $m$.
With the projected density [Eq.~\eqref{eq:qh-density}], the density link of the band $\ket{\Psi^n_{\bm k}}=\sum_mv_{n,m}(\bm k)\ket m$, with $v_n(\bm k)$ the eigenvectors of $H^{(1)}(\bm k)$, is
\begin{equation}
 \Lambda_{n,\bm q}(\bm k)=f(\bm q)\,e^{\ii\alpha_{\bm q}(\bm k)}\,
 v_n^\dagger(\bm k+\bm q)\,v_n(\bm k).
 \label{eq:chern-link}
\end{equation}
The first two factors give the uniform plaquette phase $6\pi/N_\phi$ [Eq.~\eqref{eq:qh-plaquette}], as for the square modulation [Eq.~\eqref{eq:square-plaquette-split}].
The Brillouin zone (BZ) of the magnetic cell is spanned by $\bm G_1/q$ and $\bm G_2$ and holds $N_\phi/p$ plaquettes, so for a dense momentum grid an isolated band $v_n$ has
\begin{equation}
 C=\frac{3}{p}+\frac1{2\pi}\int_{\rm BZ}\dd^2k\,F,
 \qquad F=\partial_{k_x}\mathcal V_y-\partial_{k_y}\mathcal V_x,
 \qquad \bm{\mathcal V}=\ii v_n^\dagger\nabla_{\bm k}v_n.
 \label{eq:chern-split}
\end{equation}

Let $u$ be the integer with $3qu\equiv-1\pmod p$, so that $\omega^u=e^{2\pi\ii/p}$, and let $\bm\delta_j=\bm G_j/(3q)$.
The shift $\bm k\to\bm k+\bm\delta_1$ multiplies the term $(n_1,n_2)$ of the Fourier series [Eq.~\eqref{eq:chern-fourier}] by $e^{2\pi\ii n_2/p}$, and $\bm k\to\bm k+\bm\delta_2$ multiplies it by $e^{-2\pi\ii n_1/p}$.
Since $Z^uPZ^{-u}=e^{2\pi\ii/p}P$ and $P^uZP^{-u}=e^{-2\pi\ii/p}Z$, both shifts are conjugations,
\begin{equation}
 H^{(1)}(\bm k+\bm\delta_1)=Z^uH^{(1)}(\bm k)Z^{-u},
 \qquad
 H^{(1)}(\bm k+\bm\delta_2)=P^uH^{(1)}(\bm k)P^{-u}.
 \label{eq:chern-ninefold}
\end{equation}
The energies and $F$ therefore repeat $9q$ times in the BZ; for $p=q=1$ this is the ninefold periodicity of the dispersion and of the curvature in the main text.
With $R$ the parallelogram spanned by $\bm\delta_1$ and $\bm\delta_2$, with one corner at the origin,
\begin{equation}
 C=\frac{3}{p}+9q\,\Phi,\qquad
 \Phi=\frac1{2\pi}\int_R\dd^2k\,F.
 \label{eq:chern-ninth}
\end{equation}

On opposite edges of $R$, the eigenvectors are related by the same unitaries,
\begin{equation}
 v(\bm k+\bm\delta_1)=e^{\ii\beta_1(\bm k)}Z^uv(\bm k),\qquad
 v(\bm k+\bm\delta_2)=e^{\ii\beta_2(\bm k)}P^uv(\bm k),
 \label{eq:chern-edges}
\end{equation}
so that $\bm{\mathcal V}(\bm k+\bm\delta_j)=\bm{\mathcal V}(\bm k)-\nabla_{\bm k}\beta_j(\bm k)$, and Stokes' theorem gives
\begin{equation}
 2\pi\Phi=\big[\beta_2(\bm\delta_1)-\beta_2(\bm 0)\big]
 -\big[\beta_1(\bm\delta_2)-\beta_1(\bm 0)\big].
 \label{eq:chern-stokes}
\end{equation}
Going from the origin to the corner $\bm\delta_1+\bm\delta_2$ along either pair of edges gives the same vector,
\begin{equation}
 e^{\ii[\beta_1(\bm\delta_2)+\beta_2(\bm 0)]}Z^uP^uv(\bm 0)
 =e^{\ii[\beta_2(\bm\delta_1)+\beta_1(\bm 0)]}P^uZ^uv(\bm 0).
 \label{eq:chern-corner}
\end{equation}
Since $Z^uP^u=\omega^{u^2}P^uZ^u=e^{2\pi\ii u/p}P^uZ^u$, the two phases differ by $2\pi u/p$ modulo $2\pi$, and this difference is $2\pi\Phi$ [Eq.~\eqref{eq:chern-stokes}].
Hence
\begin{equation}
 \Phi=\frac{u}p+n,\qquad n\in\mathbb Z,
 \label{eq:chern-flux}
\end{equation}
and the Chern number [Eq.~\eqref{eq:chern-ninth}] is
\begin{equation}
 C=3j+9qn,\qquad j=\frac{1+3qu}p\in\mathbb Z,\qquad pj\equiv1\pmod{3q}.
 \label{eq:chern-three}
\end{equation}
Every isolated band therefore has $C\in3\mathbb Z$, fixed modulo $9q$.
Equivalently,
\begin{equation}
 C=3t,\qquad 1=3qs+pt,\qquad s,t\in\mathbb Z,
 \label{eq:chern-diophantine}
\end{equation}
which is the Diophantine equation of magnetic bands~\cite{Thouless1982,Dana1985} at $p/(3q)$ flux quanta per cell.

The curvatures $F$ of all $p$ bands add up to zero, so their Chern numbers add up to $3$; for $p=1$ the single band has $C=3$.
For $p/q=2$, $u=1$, so $\Phi$ is a half-integer and each band has $C=3/2+9\Phi\equiv6\pmod9$.
The two Chern numbers therefore have opposite signs, and $(6,-3)$ and $(-3,6)$ are the pairs of smallest magnitude.
These constraints hold for the isolated bands of $H^{(1)}$, and for the full $H_{\rm eff}$ as long as no gap closes when the higher orders are included.

\FloatBarrier
\section{Exact diagonalization with the Coulomb interaction} \label{sec:ed-coulomb}

To test the band-folding picture beyond the quasihole-subspace projection,
we perform exact diagonalization (ED) of the Coulomb interaction in the full
many-electron Hilbert space of the Aharonov--Casher zero-mode band.
We study $N_e=11$ electrons at $N_\phi=34$ on the square torus
$\bm L_1=a(3,5)$ and $\bm L_2=a(-5,3)$, with zero magnetic
boundary twists. This geometry is commensurate with the
$\sqrt{2}\times\sqrt{2}$ modulation
\begin{equation}
\frac{B_z(\bm r)}{-B_0}
=
1+\kappa\left[
\cos\frac{2\pi x}{a}+\cos\frac{2\pi y}{a}
\right]
+\tilde\kappa\left[
\cos\frac{\pi(x+y)}{a}+\cos\frac{\pi(x-y)}{a}
\right],
\label{eq:ed-magnetic-field}
\end{equation}
where $B_0a^2=h/e$, so that
$a^2=2\pi\ell_B^2$.
We consider both weak and strong magnetic modulation, with
\begin{equation}
 (\kappa,\tilde\kappa)=
 \begin{cases}
 (-0.35,0.05),&\text{weak modulation},\\
 (-3.5,0.5),&\text{strong modulation}.
 \end{cases}
 \label{eq:ed-modulation-parameters}
\end{equation}

\begin{figure}[htbp]
 \centering
 \includegraphics[width=3.42in]{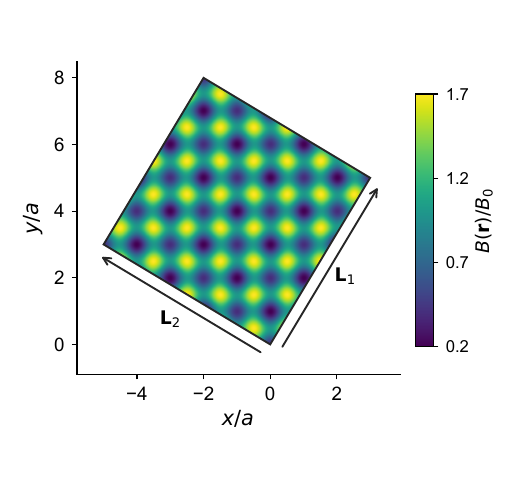}
 \caption{Magnetic-field profile $-B_z(\bm r)/B_0$ for
 $(\kappa,\tilde\kappa)=(-0.35,0.05)$.
 The outlined square is the $N_\phi=34$ torus spanned by $\bm L_1$ and
 $\bm L_2$, with opposite edges identified.}
 \label{fig:ed-real-space-geometry}
\end{figure}

The primitive translations of the doubled modulation cell are
$\tilde{\bm a}_1=(a,a)$ and $\tilde{\bm a}_2=(-a,a)$.
The torus periods satisfy
\begin{equation}
 \bm L_1=4\tilde{\bm a}_1+\tilde{\bm a}_2,\qquad
 \bm L_2=-\tilde{\bm a}_1+4\tilde{\bm a}_2,
 \qquad |\bm L_1\times\bm L_2|=34a^2,
 \label{eq:ed-torus-geometry}
\end{equation}
so the torus contains 34 original square cells, or 17 doubled cells.
Figure~\ref{fig:ed-real-space-geometry} shows the weak-modulation
magnetic-field profile on this torus.

The Coulomb Hamiltonian projected onto the Aharonov--Casher
zero-mode band is
\begin{equation}
\frac{H_{\mathrm C}}{E_C}
=
\frac{1}{2N_\phi}
\sum_{\bm q\neq0}
\frac{
\mathcal P_{\rm AC}:\rho_{\bm q}\rho_{-\bm q}:\mathcal P_{\rm AC}
}{
|\bm q|\ell_B
},
\qquad
E_C=\frac{e^2}{\epsilon\ell_B},
\end{equation}
where $\rho_{\bm q}=\sum_j e^{\ii\bm q\cdot\bm r_j}$,
$\mathcal P_{\rm AC}$ denotes the band projection, and $\bm q$ runs over
the torus reciprocal lattice. Normal ordering removes
self-interactions, while omitting $\bm q=0$ implements a uniform
neutralizing background. We obtain the four lowest energies
in each of the 17 folded momentum sectors using Lanczos
diagonalization.

\begin{figure}[t]
    \centering
    \includegraphics[width=\linewidth]{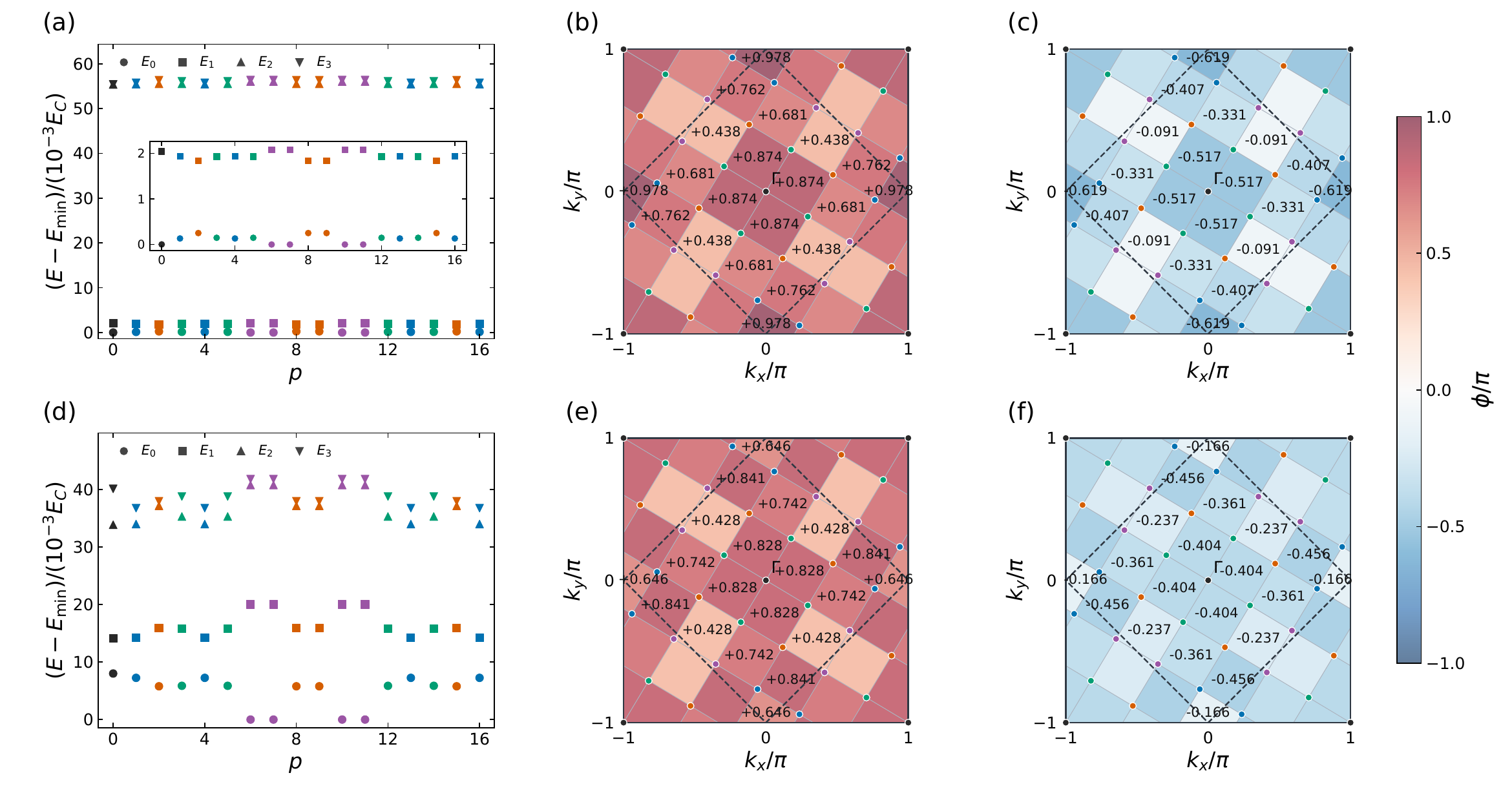}
    \caption{
    Quasihole bands with opposite Chern numbers from exact diagonalization of the
    Coulomb interaction at $(N_\phi,N_e)=(34,11)$.
    The top and bottom rows show the weak and strong modulation cases
    in Eq.~\eqref{eq:ed-modulation-parameters}, respectively.
    (a,d) The four lowest energies in each folded momentum sector $p$,
    relative to the lowest energy $E_{\min}$.
    The inset in (a) resolves the two lowest bands.
    (b,e) and (c,f) Plaquette phases $\phi/\pi$ for the lower and upper
    bands, respectively, with $\phi=\gamma_\square$ defined in
    Eq.~\eqref{eq:curvature-def}.
    Both parameter sets give $(C_-,C_+)=(6,-3)$.
    Dashed diamonds mark the folded Brillouin zone;
    matching marker colors indicate sectors related by fourfold rotation.
    }
    \label{fig:EDcoulomb}
\end{figure}

Figure~\ref{fig:EDcoulomb} shows two isolated low-energy quasihole
bands at weak modulation, consistent with the modulation
picture for $\alpha V_{\mathrm{TK}}+V$.
At stronger modulation, their gap to higher many-body states
becomes comparable to the splitting between the two bands.
We evaluate the Berry curvature and Chern numbers from the density links
using Eqs.~\eqref{eq:density-link} and \eqref{eq:curvature-def}, with
positively oriented plaquettes on the reciprocal mesh of the tilted torus.
For both parameter sets, the lower and upper bands have
$(C_-,C_+)=(6,-3)$.
These nonperturbative Coulomb results are consistent with the
opposite-Chern Laughlin quasihole bands in the
weak-modulation regime.

\FloatBarrier

\section{Monte Carlo method} \label{sec:numerical-simulation}

\subsection{Matrix elements and importance sampling}

We evaluate the projected problem with the full envelope $\Gamma$, without expanding in the modulation.
Writing $\Phi_{\bm k\alpha}(\bm z)=\langle\bm z|\Psi_{\bm k,\alpha}\rangle$ and $\dd\bm z=\prod_i\dd^2r_i$, the required matrices are
\begin{equation}
 \begin{aligned}
 S_{\alpha\beta}(\bm k)&=\int\dd\bm z\,
 \Phi_{\bm k\alpha}^*\Gamma^2\Phi_{\bm k\beta},\\
 H_{\alpha\beta}(\bm k)&=\int\dd\bm z\,
 \Phi_{\bm k\alpha}^*\Gamma^2 V\Phi_{\bm k\beta},\\
 [D_{\bm q}(\bm k)]_{\alpha\beta}&=\int\dd\bm z\,
 \Phi_{\bm k+\bm q,\alpha}^*\Gamma^2\rho_{\bm q}\Phi_{\bm k\beta},
 \qquad \rho_{\bm q}(\bm z)=\sum_i e^{\ii\bm q\cdot\bm r_i}.
 \end{aligned}
 \label{eq:mc-matrices}
\end{equation}
There is one component per momentum for the single band and two for the folded bands.
The smallest transfers $\bm q_x=(2\pi/L_x)\hat{\bm x}$ and $\bm q_y=(2\pi/L_y)\hat{\bm y}$ give the density links used for the curvature.
At a zone boundary the endpoint is folded back into the stored basis, while the density operator retains the small transfer $\bm q$.

The Metropolis Monte Carlo (MC) distribution is a positive mixture of uniform-field localized quasihole states $\psi_j(\bm z)=\langle\bm z|\psi^{\sigma_j}_{\eta_j}\rangle$ spanning the quasihole space:
\begin{equation}
 \mu(\bm z)=\sum_j|\psi_j(\bm z)|^2,\qquad
 p_0(\bm z)=\frac{\mu(\bm z)}{\mathcal Z_0},\qquad
 \mathcal Z_0=\int\dd\bm z\,\mu(\bm z).
 \label{eq:mc-distribution}
\end{equation}
On each configuration, the Bloch amplitudes are evaluated as linear combinations of the same $\psi_j$.
For any integrand $F$ of these matrix elements [Eq.~\eqref{eq:mc-matrices}], the estimator is
\begin{equation}
 \frac1{\mathcal Z_0}\int\dd\bm z\,F(\bm z)
 =\left\langle\frac{F}{\mu}\right\rangle_{p_0}
 \simeq\frac1{N_{\rm samp}}\sum_{s=1}^{N_{\rm samp}}
 \frac{F(\bm z_s)}{\mu(\bm z_s)}.
 \label{eq:mc-estimator}
\end{equation}
The common normalization is fixed by the uniform-state overlap; it cancels from normalized observables.
The envelope $\Gamma^2$ enters the estimators, rather than the sampling distribution.
All entries of $S,H,D$ are accumulated on the same configurations, retaining their statistical correlations.
One sweep attempts a displacement of each electron, with periodic wrapping; a collective translation of all electrons is additionally proposed every ten sweeps.
Each chain has an independent random seed and, unless stated otherwise, $2048$ warmup sweeps.

The interaction on a torus of area $A$ is
\begin{equation}
 V=\frac1{2A}\sum_{\bm q\ne0}\frac{2\pi e^2}{\epsilon q}\tanh(qd)
 (\rho_{\bm q}\rho_{-\bm q}-N_e),\qquad
 E_C=\frac{e^2}{\epsilon\ell_B},
 \label{eq:square-screened-interaction}
\end{equation}
with torus-compatible $\bm q$.
To reduce the noise in small energy differences, we use the exact uniform-field identities $S^{(0)}=I$ and $H^{(0)}=E_0I$ and sample
\begin{equation}
 \begin{aligned}
 S-I&=\bra{\Psi}(\Gamma^2-1)\ket{\Psi},\\
 H-E_0S&=\bra{\Psi}(V-E_0)(\Gamma^2-1)\ket{\Psi},
 \end{aligned}
 \label{eq:mc-uniform-subtraction}
\end{equation}
where momentum and component indices are suppressed and $E_0$ is estimated from the same chains.
The static estimators also average over symmetry-related magnetic translations.
We solve the generalized eigenvalue problem [Eq.~\eqref{eq:ac-projected-problem}], and form the density links $c_n^\dagger(\bm k+\bm q)D_{\bm q}(\bm k)c_n(\bm k)$.
Their plaquette phases give the Berry curvature and lattice Chern numbers as in the main text.

For the periodic electric potential, we set $\Gamma=1$ and replace $V$ by $\hat V_{\rm el}$ in Eq.~\eqref{eq:mc-matrices}.
We integrate the electric-potential matrix and the density links directly in the complete space $\mathcal H_{\rm qh}$, without imposing the form-factor parametrization of Eq.~\eqref{eq:electric-projection} on the sampled matrices.
The uniform-state metric is exactly proportional to the identity; its common normalization is estimated from the trace of the sampled overlap matrix and recomputed in every bootstrap draw.
These calculations retain the projection onto $\mathcal H_{\rm qh}$ and exclude mixing with excited states of the parent Hamiltonian.

\subsection{Error estimates}

We resample whole independent chains with replacement, using the same chain indices for $H$, $S$, and $D$ at all momenta.
Each bootstrap draw recomputes the uniform normalization, $E_0$, the generalized eigenstates, and the density links.
Reported errors are one standard deviation of the resulting observable across the draws.
Thus correlations within a chain and between the matrices are retained; consecutive configurations are not treated as independent samples.
The Chern-number stability is assessed from its distribution over these draws, together with checks at larger system size.

Figures~1 and 3 of the main text use $256$ and $32$ independent chains, respectively, with $32768$ retained configurations per chain.
The corresponding error estimates use $400$ bootstrap draws each.
The single-band coefficients $\chi_1,\xi_1$ are measured in the uniform state.
For the folded square bands, $\chi_1,\tilde\chi$ are measured using $512$ uniform-state chains of $65536$ configurations at $N_\phi=100$, with $2000$ bootstrap draws.

The triangular calculation in Fig.~\ref{fig:tri-c6-numerics} uses $32$ chains of $16384$ configurations and $200$ bootstrap draws.

For the periodic electric potential, we extract the uniform form factors by the trace identity that follows from the projected density [Eq.~\eqref{eq:qh-density}],
\begin{equation}
 f(\bm q)=\frac1{N_\phi}\operatorname{Tr}_{\mathcal H_{\rm qh}}
 \bigl[U_{\bm d(\bm q)}^\dagger\mathcal P\rho_{\bm q}\mathcal P\bigr].
 \label{eq:mc-uniform-form-factor}
\end{equation}
This form-factor calculation is performed at $(N_\phi,N_e)=(100,33)$ using $32$ chains of $32768$ configurations and $200$ whole-chain bootstrap draws.
We average over the four symmetry-related wave vectors in each harmonic and recompute $f_1$ and $f_2$ in every draw.
The direct calculation in Fig.~\ref{fig:electric-periodic-bands} uses $32$ independent chains with approximately $10^6$ retained configurations in total.
The energy and curvature uncertainties describe sampling errors; finite-size and finite-mesh effects are not included.

\subsection{Wave-packet dynamics}

For Fig.~2 of the main text, we use $(N_\phi,N_e)=(256,85)$ on the square torus $L=16a$, with $(\kappa,\tilde\kappa)=(-0.35,0.05)$ and $d=\ell_B$.
The generalized eigenvalue problem [Eq.~\eqref{eq:ac-projected-problem}] gives the orthonormal band states $\ket{\Psi^n_{\bm k}}=\Gamma\sum_\alpha c_{n,\alpha}(\bm k)\ket{\Psi_{\bm k,\alpha}}$, with $n=+$ for the upper band ($C=-3$) and $n=-$ for the lower band ($C=6$).
We expand the evolving state in these states and start from the quasihole localized at the origin and projected onto band $n_0$,
\begin{equation}
 \ket{\zeta(t)}=\sum_{\bm k,n}b_{\bm k n}(t)\ket{\Psi^n_{\bm k}},
 \qquad
 b_{\bm k n}(0)\propto\delta_{n n_0}
 \bra{\Psi^{n_0}_{\bm k}}\Gamma\ket{\psi^\sigma_0},
 \label{eq:mc-dynamics-coefficients}
\end{equation}
with $\sum_{\bm k,n}|b_{\bm k n}(0)|^2=1$.

With $\bm q_d=q(\hat{\bm x}+\hat{\bm y})$ and $q=2\pi/L$, the potential energy of the electrons used in the main text is
\begin{equation}
 \hat V_{\mathcal E}(t)=\frac{e\mathcal E(t)}{\sqrt2q}\sum_i\sin(\bm q_d\cdot\bm r_i)
 =\frac{e\mathcal E(t)}{2\ii\sqrt2q}\big(\rho_{\bm q_d}-\rho_{\bm q_d}^\dagger\big).
 \label{eq:mc-drive}
\end{equation}
Its matrix $V_{\mathcal E}^{\rm proj}(t)$ between the band states therefore follows from that of $\rho_{\bm q_d}$, which raises the momentum by $\bm q_d$,
\begin{equation}
 \bra{\Psi^m_{\bm k+\bm q_d}}\rho_{\bm q_d}\ket{\Psi^n_{\bm k}}
 =c_m^\dagger(\bm k+\bm q_d)\,D_{\bm q_d}(\bm k)\,c_n(\bm k),
 \label{eq:mc-dynamics-density-matrix}
\end{equation}
where $D_{\bm q}(\bm k)$ is the sampled matrix of $\Gamma^2\rho_{\bm q}$ between the Bloch components [Eq.~\eqref{eq:mc-matrices}] and $\bm k+\bm q_d$ is folded into the reduced zone.
We sample $D_{\bm q_d}$ with $256$ chains of $32768$ configurations.
The field is switched on as $\mathcal E(t)=\mathcal E g(\tau)$, with
\begin{equation}
 \tau=\frac{e\mathcal E\ell_Bt}{\hbar},\qquad
 \frac{e\mathcal E\ell_B}{E_C}=4.51\times10^{-5},\qquad
 g(\tau)=
 \begin{cases}
 \tfrac12[1-\cos(\pi\tau/\tau_r)],&0\leq\tau<\tau_r,\\
 1,&\tau\geq\tau_r,
 \end{cases}
 \qquad \tau_r=0.588,
 \label{eq:mc-dynamics-ramp}
\end{equation}
and the amplitudes obey
\begin{equation}
 \ii\hbar\frac{\dd\bm b}{\dd t}
 =\left[\operatorname{diag}_{\bm k,n}E_n(\bm k)
 +V_{\mathcal E}^{\rm proj}(t)\right]\bm b.
 \label{eq:mc-dynamics-schrodinger}
\end{equation}
We integrate this equation during the ramp and, after it, propagate by diagonalizing the constant Hamiltonian.
The dispersions of both bands and the couplings between them are kept throughout; only the initial packet is projected onto one band.

To obtain the density, we return to the Bloch components, $a_{\bm k\alpha}(t)=\sum_nc_{n,\alpha}(\bm k)\,b_{\bm k n}(t)$, so that $\ket{\zeta(t)}$ has the wave function $\Upsilon_t(\bm z)=\Gamma(\bm z)\sum_{\bm k,\alpha}a_{\bm k\alpha}(t)\Phi_{\bm k\alpha}(\bm z)$.
Reweighting the same mixture $p_0$ [Eq.~\eqref{eq:mc-distribution}] gives
\begin{equation}
 \lambda_t(\bm z)=\frac{|\Upsilon_t(\bm z)|^2}{\mu(\bm z)},\qquad
 \langle\rho_{\bm q}\rangle_t
 =\frac{\langle\lambda_t\rho_{\bm q}\rangle_{p_0}}
 {\langle\lambda_t\rangle_{p_0}},\qquad
 n(\bm r,t)=\frac1A\sum_{\bm q}e^{-\ii\bm q\cdot\bm r}
 \langle\rho_{\bm q}\rangle_t,
 \label{eq:mc-density}
\end{equation}
where the numerator and the denominator are summed over chains before their ratio is taken, and $\langle\rho_{\bm0}\rangle_t=N_e$.
We use $192$ chains of $16384$ configurations for both bands at $\tau=0,0.1,\ldots,2$, with the cutoff $|\bm q|\ell_B\leq4$.
With the density $n_{\rm bg}$ without a quasihole, sampled at $(N_\phi,N_e)=(324,108)$ and continued periodically in the same modulation, $\delta n=n-n_{\rm bg}$ integrates to $-1/3$.

The displacements parallel and perpendicular to the field, along $(\hat{\bm x}\pm\hat{\bm y})/\sqrt2$, follow from the phases of the periodic density moments,
\begin{equation}
 M_\pm(t)=-3\int\dd^2r\,\delta n(\bm r,t)e^{\ii q(x\pm y)},\qquad
 \Delta R_{\parallel,\perp}(t)=\frac{
 \operatorname{Arg}[M_\pm(t)M_\pm(0)^*]}{\sqrt2q}.
 \label{eq:square-dynamics-centroid}
\end{equation}
For a deficit localized well within the torus, this estimator approaches the charge center, and $n_{\rm bg}$, which has no component at $q(1,\pm1)$, drops out of $M_\pm$.
Error bars for the two trajectories use $128$ whole-chain bootstrap resamples shared across states and times, holding the Hamiltonian, overlap, drive, and initial state fixed.
For each draw we recompute the density ratios, the moment phases, and the displacement relative to the same draw at $t=0$.

\bibliography{references,main_refs}